\documentclass[a4paper,11pt]{article}
\usepackage{jheppub} 
\usepackage{xcolor}
\usepackage{tikz}
\usetikzlibrary{math}
\usepackage{amsmath}
\usepackage{amsmath,amssymb}
\usepackage{mathtools}

\DeclarePairedDelimiterX\braket[2]{\langle}{\rangle}{#1\,\delimsize\vert\,\mathopen{}#2}

\title{Holographic complexity of conformal fields in global de Sitter spacetime}

\author[]{Sanhita Parihar}
\author[]{and Shubho R. Roy}

\affiliation[]{Indian Institute of Technology Hyderabad,Kandi, Sangareddy, Telengana 502285, India}

\emailAdd{sanhita.hepth@gmail.com, roy.shubho@gmail.com}

\abstract{We compute the holographic complexity of conformal quantum fields in rigid global de Sitter spacetime (dS$_{d}$) using the volume and action prescriptions. First we consider AdS$_{d+1}$ spacetime in global dS$_{d}$ foliations, and compute the complexity of the CFT supported on the global dS$_{d}$ conformal boundary. Next, we consider CFT supported on a global dS$_d$ (UV) brane embedded in AdS$_{d+1}$ spacetime, and compute the holographic complexity in this brane set up. We compare and contrast the results in the two cases, as well as with related results in the literature obtained in alternative holographic set ups involving patches of de Sitter spacetime covered by static coordinates or conformal (Poincar\'e) coordinates.}

\begin{document}
\maketitle
\flushbottom

\section{Introduction \& Summary} 

Studying quantum field theory in a curved spacetime is an essential step on the way towards understanding quantum gravity. As there are substantial evidence that our universe is curved, and undergoing as accelerating expansion with a very small yet positive cosmological constant\cite{SupernovaCosmologyProject:1996grv,SupernovaSearchTeam:1998fmf, Suzuki_2012}. As such, de Sitter spacetime provides a  model of our universe at late times. De Sitter spacetime also serves as a useful toy model for studying the early universe, in particular for the inflationary era. In either case, the (positive) curvatures involved are much smaller that the Planck scale. As such, studying quantum field theory in a de Sitter background is expected to not just be insightful, but to play an important role in the physics of the early and late universe. A general study of QFT in de Sitter spacetime is quite challenging due to several factors. First of all the global de Sitter spacetime does not admit a global timelike killing vector. In other words de Sitter spacetime does not have time translation symmetry and therefore there is no conservation of energy: Hamiltonian is not defined
properly and the quantization procedure does not proceed smoothly: creation-annihilation operators and the associated notion of particles becomes time-dependent. To circumvent this, one can resort to working in static coordinates, whereby the metric is static and admits a timelike Killing vector. Thus a notion of conserved energy and time-independent creation-annihilation operators exists. However the static patch covers only a part of full de Sitter spacetime and is bounded by horizons which then create issues with unitary: information may disappear outside the static patch (horizons) and thus defining a S-matrix becomes untenable. Another salient feature of field theory in de Sitter spacetime is the nonuniqueness of the vacua: there exists of a family of inequivalent vacua with de Sitter isometries ($\alpha$ vacua) \cite{Chernikov:1968zm, Mottola:1984ar, Allen:1985ux}. It is a common practice to sidestep this issue, and instead work with Bunch-Davies vacuum which respects symmetries of $(d+1)$ dimensional de Sitter spacetime $SO(1,d+1)$ obtained from Euclidean de Sitter (sphere) via analytic continuation \cite{Chernikov:1968zm,Schomblond:1976xc}. This prescription is also followed for the additional advantage that at short spacetime scales (much shorter in comparison to the de Sitter radius) the Bunch-Davies vacuum corresponds to the standard QFT vacuum in Minkowski spacetime. A key challenging feature of interacting QFT on de Sitter spacetime is ubiquitous existence of infrared (IR) divergences. It is well known that if we consider a massless scalar field in de Sitter then the two point correlation function is not invariant under de Sitter symmetries, it grows linearly with time \cite{Linde:1982uu,Vilenkin:1982wt}, which strongly points towards the instability of de Sitter space in presence of quantum fields \cite{Polarski:1991ek}. Similar IR divergences has been observed for massless interacting scalar fields, and scalar fields coupled with gauge fields \cite{Ford:1984hs}. Furthermore in presence of gravitational fluctuations the graviton loops also exhibit IR divergence \cite{Tsamis:1996qq}. There have been multiple proposals to address or circumvent these IR divergences in de Sitter spacetime. These include stochastic approach in which the problem is mapped to an classical stochastic process \cite{Starobinsky:1994bd}, nonperturbative methods \cite{Burgess:2010dd,Moreau:2018lmz},and approach based on refined definition of observables and gauge-invariant quantities \cite{Higuchi:2011vw,Bernar:2014lna}. Despise these efforts a fully unified and universally accepted resolution of IR divergences in de Sitter remains an open issue. For a more detailed discussion on IR divergences, instability of de Sitter and its possible resolution refer to the review \cite{Hu:2018nxy}. In this paper we focus our attention to strongly interacting conformal fields in de Sitter spacetime, via holography, i.e. using a gravitational AdS dual, and in particular employing a specific quantum information tool which has come into prominence in the holographic context, namely quantum computational complexity.\\ 

For a generic quantum system the \textit{quantum computational complexity} is defined as minimum number of simple operations needed to reach a target state from a reference state in the Hilbert space of the quantum system. In quantum information theory the simple operations are defined in terms of \textit{gates}. While this definition is perfect for discrete quantum system its generalization to continuous quantum systems such as quantum field theory is not well established. There are several proposals for complexity of an continuous system\cite{Nielsen:2006cea,Jefferson:2017sdb, Chagnet:2021uvi} but there is no universal definition of quantum computational complexity in continuum field theory. However we will not pursue this direction further but instead we work with holographic dual of conformal field theory living on global de Sitter space as this will allow us to capture non-perturbative/strong coupling description of quantum field theory. \\

The foundation and motivation for this is rooted in two well celebrated conjectures. The first is AdS/CFT correspondence \cite{Maldacena:1997re, Witten:1998qj} which provide us a framework to probe the strongly coupled regime of quantum field theories through dual description in terms of an higher dimensional classical gravity theory. The second is Ryu-Takanayagi proposal of holographic entanglement entropy\cite{Ryu:2006bv,Ryu:2006ef} which identifies the area of an extremal co-dimension two surface in bulk with the entanglement entropy of the boundary theory. As the entanglement entropy of a quantum field theory is understood as an measure of quantum information, this proposal associates how the information of bulk geometry is encoded in the boundary degree of freedom. This answers an central and enduring question in holography that is how the bulk spacetime geometry and its dynamics is encoded in the boundary quantum field theory. This geometric realization of entanglement provided a concrete link between spacetime geometry and quantum information, and naturally led to increased interest in other quantum information–theoretic probes that might further illuminate the emergence and structure of the bulk. One such probe is quantum computational complexity of the boundary quantum field theory, which enable to understand the growth of Einstein-Rosen bridge even after the thermalization which entanglement entropy fails to capture \cite{Susskind:2014moa}. As a result, quantum complexity offers a complementary and more sensitive diagnostic of bulk interior dynamics, capturing aspects of spacetime evolution that are invisible to entanglement entropy alone.\\ 

Nowadays there are many different proposals for the bulk dual of the quantum computational complexity of boundary field theory \cite{Belin:2021bga,Belin:2022xmt,Myers:2024vve}. It has been well accepted that there are multiple co-dimension one and co-dimension zero gravitational observables in the bulk which satisfy the properties of quantum computation complexity of boundary field theory; such as late time linear growth and switch-back effects. Two of these observables which were initially conjectured dual to the quantum computational complexity are, maximal volume of an spacelike hypersurface known as the volume conjecture \cite{Susskind:2014rva}, and the on-shell action on Wheeler-De-Witt patch (a union of all spacelike hypersurfaces anchored to the boundary) this conjecture is known as Action conjecture \cite{Brown:2015lvg,Brown:2015bva}.\\

The study of quantum computational complexity in quantum field theories with holographic duals in asymptotically AdS spacetime has been explored extensively in past years \cite{Carmi:2016wjl, Carmi:2017jqz, Chapman:2016hwi, Alishahiha:2015rta}. Motivated by these developments, a natural and important question is how to define and characterize computational complexity in field theories whose gravitational duals are asymptotically de Sitter (dS) spacetime rather than AdS. This problem is substantially more subtle, reflecting the fact that holography in de Sitter space itself is not yet fully understood. Several approaches to de Sitter holography have been proposed, each emphasizing different aspects of the spacetime. One of the earliest proposals is the dS/CFT correspondence, originally suggested by Strominger \cite{Strominger:2001pn}, in which quantum gravity in $(d+1)$-dimensional de Sitter space is conjectured to be dual to a Euclidean conformal field theory living at future (or past) infinity. This conjecture is supported by calculating the two point correlation function of a massive scalar field with end points at past infinity, which matches the form of CFT correlator on plane, with complex conformal weight if mass $m^{2}\ell_{dS}^{2}>\frac{d^2}{4}$.  While conceptually appealing, the Euclidean and non-unitary nature of the putative dual conformal field theory complicates the interpretation of dynamical observables. Moreover, infrared (IR) divergences in de Sitter quantum field theory particularly for light or massless fields have been argued to reflect deeper pathologies of the dual description\cite{Banks:2005tn}.\\

A second line of thought focuses on the static patch of de Sitter spacetime. Unlike global de Sitter coordinates, the static patch metric is time-independent and admits a timelike killing vector, making it more amenable to notions of equilibrium and thermodynamics. However, the static patch covers only the causal region accessible to a single observer and is bounded by an observer-dependent cosmological horizon with an associated Gibbons–Hawking temperature and entropy. The corresponding holographic proposal often referred to as static patch holography, suggests that the boundary CFT lives on the stretched (cosmological) horizon \cite{Balasubramanian:2002zh,Susskind:2021omt}. In this setting, holographic complexity has been studied \cite{Jorstad:2022mls} following a modified volume and action conjectures. The complexity of the boundary field theory living on cosmological horizon has been found to show hyperfast growth \emph{i.e.} the complexity becomes exponentially large in some finite boundary time, this behavior is consistent with earlier expectations and arguments in de Sitter space \cite{Susskind:2021esx} that holographic complexity in de Sitter exhibit hyperfast growth due to the inflationary (exponential) growth of interior of de Sitter.\\

A third proposal is the dS/dS correspondence \cite{Alishahiha:2004md}, which posits that gravity in $(d+1)$-dimensional static de Sitter space is dual to two copies of $d$-dimensional conformal field theory coupled to gravity living on the lower-dimensional de Sitter spacetime dS$_d$. In this proposal there has been an significant development \cite{Gorbenko:2018oov,Lewkowycz:2019xse,Coleman:2021nor,Batra:2024kjl} in which the holographic dual of the warped throat on the gravity side of dS/dS duality is constructed by deforming CFT dual to AdS geometry with a specific combination of $T\bar{T}$ and cosmological constant $\Lambda$ deformation. These works form a robust foundation to understand dS/dS correspondence starting from AdS holography. In this setup the entanglement entropy is studied and agreement is found between bulk and boundary calculations \cite{Gorbenko:2018oov}. The study holographic complexity in this setup remains an open problem.\\

In this paper, our objective is somewhat different from the standard holographic studies of computational complexity, which typically focus on boundary field theories defined on static or asymptotically flat backgrounds and the quantum computational complexity is studied via there gravity duals. Instead, we aim to directly study the quantum computational complexity of a boundary field theory living on a de Sitter spacetime. Investigating complexity in a field theory formulated on de Sitter space is interesting and important for several reasons. First, it provides a concrete example of quantum complexity in a genuinely time-dependent background. In particular, when written in global coordinates, the de Sitter metric exhibits explicit time dependence, implying that the Hamiltonian of the theory itself is time dependent. This feature makes de Sitter space a natural laboratory for understanding how complexity evolves in non-stationary quantum systems.
\begin{eqnarray}
ds^{2} = dr^{2} + \sinh^{2} r \left( -dt^{2} + \cosh^{2} t\, d\Omega_{d-1}^{2} \right) .
\end{eqnarray}
The explicit time dependence of the background geometry introduces new conceptual and technical challenges in defining and computing complexity, particularly in contrast to the well-studied AdS case where the boundary geometry is typically static. Understanding how complexity responds to the expansion of space and to the cosmological time evolution may shed light on the interplay between quantum information measures and cosmological dynamics.\\

Furthermore, quantum computational complexity in continuum field theory is an UV divergent quantity this is physically intuitive as complexity is defined as the minimum number of operations required to create a target state, achieving higher precision demands more operations, leading naturally to divergences of complexity. Therefore the dependence of complexity on a UV cutoff is an intrinsic and meaningful aspect of the framework. However quantum gravity in de Sitter spacetime is known to be plagued by infrared divergences \cite{Linde:1982uu,Ford:1984hs,Tsamis:1996qq}. Studying quantum information–theoretic quantities, such as quantum computational complexity, in this setting provides an opportunity to probe not only its familiar UV structure but also its response to long-distance, IR effects. In particular, analyzing how complexity behaves in the presence of IR divergences may offer insight into their physical significance and into the robustness of quantum information measures in cosmological spacetime. From this perspective, complexity serves not only as a diagnostic of quantum dynamics on curved backgrounds but also as a potential tool for exploring the distinctive infrared structure of de Sitter space.\\

A similar setup was considered in \cite{Hawking:2000da} where the conformal field theory coupled to the gravity theory on de-Sitter spacetime. This has been achieved by inserting a two sided brane\footnote{So the two patches of AdS are glued at brane} with non-zero tension $T$ at some radius $r=r_{B}$ in three dimensional AdS spacetime, and the induced geometry on the brane is de Sitter. The action of the three dimensional gravity theory now becomes,
\begin{align}
    S=\frac{1}{16\pi G}\int d^{3}x\sqrt{
    g}
    \left(R+\frac{2}{L^2}\right)+T\int d^{2}\sigma \sqrt{h}
\end{align}
In this setup an agreement between the entropy calculated by three dimensional AdS bulk geometry and two dimensional induced de Sitter geometry on brane is observed and later this has been given an microscopic interpretation from the entanglement entropy across the de Sitter horizon.
But the de Sitter geometry was described in static coordinates and therefore lacks the global aspects of de Sitter spacetime such as explicit time dependence. Still this result provides an explicit realization of the idea that the de Sitter horizon entropy can be understood as entanglement entropy of the underlying quantum degrees of freedom. This setup widely known as braneworld holography or double holography has much wider understanding \cite{Randall:1999vf,Karch:2000ct} and has been applied in a wide range of context \cite{Emparan:2006ni,Emparan:2022ijy} including geometric interpretation of the island proposal of entanglement entropy \cite{Almheiri:2019hni,Chen:2020uac}, and investigation of  quantum aspects of gravity and blackholes \cite{Emparan:1999fd,Emparan:1999wa,Emparan:2002px}. While in \cite{Hawking:2000da} the authors commented on possible extensions to higher dimensions, their analysis did not include explicit computations in those cases. A detailed higher-dimensional treatment was subsequently provided in \cite{Iwashita:2006zj}. There it was found that the entanglement entropy of the CFT matches the entropy associated with the cosmological horizon namely, one quarter of the horizon area (computed from the brane perspective) only in a specific limit in which the brane is placed very close to the AdS boundary. In contrast, the entanglement entropy was shown to exactly agree with the entropy obtained from the bulk partition function, highlighting an important distinction between the bulk and induced gravitational descriptions. These results were further generalized to higher-derivative theories of gravity, including Gauss–Bonnet and Lovelock theories, e.g. in \cite{Kushihara:2021fbr}.\\

Despite these advances, several important questions remain open. In particular, it is not yet clear how the presence of black holes in the bulk spacetime modifies the relationship between bulk entropy, brane entropy, and the entanglement entropy of the boundary theory. Furthermore, to the best of our knowledge, a systematic and explicit investigation of quantum computational complexity within this braneworld de Sitter framework is still lacking. To address this gap we first study complexity in the de Sitter slicing of AdS spacetime, which provides a controlled setting for understanding the relevant features before introducing a de Sitter brane. Building on this analysis, we also comment on the implications of incorporating a de Sitter brane with non-zero tension into the setup. A further generalization would be to account for Einstein-Hilbert term in the brane action which we leave as an future goal.\\

The paper is organized as follows, in section \ref{sec: setup} we provide the details of our setup. Then in section \ref{sec: Volume} we study holographic complexity in de Sitter space by following the volume conjecture. We find that the explicit time dependence of the global de Sitter constraint the computation of volume complexity for general d-dimensional de Sitter spacetime and we could not be found an analytic expression in a general setting when the spacelike hypersurface is anchored at some finite boundary time $t=t_{\star}$. Therefore we rely on numerical analysis which is indeed insightful. From the numerical behavior of volume complexity we found the it depends on the UV cutoff  and boundary time exponentially $~e^{(d-1)\Lambda}e^{(d-1)t_{\star}}$. This result indicates towards the extrinsic nature of complexity as the complexity exhibit proportionality to spatial volume $~(\sinh{\Lambda}\cosh{t_{\star}})^{d-1}$. We also confirm this dependence on UV cutoff in the particular case when the boundary anchoring time $t_{\star}=0$ and find the analytic expression for complexity. Later in section \ref{sec: action} we study the complexity following action conjecture. This computation of on-shell action over the WDW patch can be done analytically in full generality for any d-dimensional de Sitter spacetime and boundary time $t_{\star}$. Although the final expression of complexity looks intricate, we highlight some key features and compare them to the complexity following from volume conjecture. Notably, the leading behavior of complexity following the action conjecture is same as that we have found following the volume conjecture \emph{i.e.} complexity$~e^{(d-1)\Lambda}e^{(d-1)t_{\star}}$. Another distinguishing aspect of the action based complexity is the presence of logarithmic divergence $~\log{\epsilon}$ in odd dimensional de Sitter spacetime where $\epsilon$ is the lattice spacing. Furthermore in section \ref{sec: brane} we discuss insertion of a brane with non vanishing tension in bulk AdS space, which induces an dynamical theory of gravity on the brane with de Sitter geometry on the brane. This setup induces an coupling between the CFT living over de geometry to the intrinsic gravity theory on the brane. Then we proceed to study of complexity in this setup following both action and volume conjecture. The resulting volume and action complexity have the exact same qualitative features that we found in absence of brane. The only quantitative difference is exhibited in the increment (doubling) of complexity. This increment can be traced to the doubling of spacetime due to gluing of two copies of AdS spacetime at the brane. Therefore our analysis indicates that insertion of a tensional brane do not create any difference to the quantum entanglement of degrees of freedom. But it certainly give an new interpretation to the complexity from the brane perspective \emph{i.e.} from the perspective of an CFT coupled to an higher derivative theory of gravity that is induced on the brane. We expect to see a non-trivial change in complexity if we include Einstein-Hilbert action corresponding the brane which we have not considered in this work. We conclude and highlight our major finding with a discussion of some outlooks in section \ref{sec:conclusion}.

\section{AdS spacetime in global dS foliations}\label{sec: setup}
To study the complexity of a CFT in global de Sitter spacetime, we consider a $(d+1)$-dimensional AdS spacetime foliated with global de Sitter slices,
\begin{eqnarray}
      ds^{2}=dr^{2}+\sinh^{2}{r}(-dt^{2}+\cosh^{2}{t}~d\Omega_{d-1}^{2})  
\end{eqnarray}
 In this parametrization $(d+1)$-dimensional AdS spacetime metric is recast as a fibration over $d$-dimensional global de Sitter fibers with warp factor of $\sinh{r}$ and  both the AdS and dS radius set to unity. This parametrization can be realized from the embedding perspective of $\text{AdS}_{d+1}$ hyperboloid in $\mathbb{R}^{2,d}$ in which each $\text{dS}_{d}$ slice is an codimension one hyperboloid embbeded in $\mathbb{R}^{d,1}$. The explicit relation between standard global AdS coordinates and the coordinates employed in de Sitter slicing are explicitly given in appendix \ref{sec: ads_ds_coord}.

\begin{figure}
    \centering
\begin{tikzpicture}[scale=4]

\fill[blue!15] (0,0) -- (0.5,0.5) -- (0,1) -- cycle;
\fill[blue!15] (1,0) -- (0.5,0.5) -- (1,1) -- cycle;

\draw[thick] (0,0) -- (0,1);
\draw[thick] (0,0) -- (1,0);
\draw[thick] (0,1) -- (1,1);
\draw[thick] (1,0) -- (1,1);
\draw[thick] (0,0) -- (0.5,0.5);
\draw[thick] (0,1) -- (0.5,0.5);
\draw[thick] (1,1) -- (0.5,0.5);
\draw[thick] (1,0) -- (0.5,0.5);
\draw[dashed] (0.5,0) -- (0.5,1);

\node[rotate=90] at (-0.05,0.5) { $r=\infty$};
\node[rotate=45] at (0.3,0.24) { $r=0$};
\node[rotate=-45] at (0.3,0.8) { $r=0$};
\node at (-0.1,-0.05) { $\tau=0$};
\node at (-0.1,1.05) { $\tau=\pi$};

\end{tikzpicture}

\caption{A schematic diagram of the region of AdS spacetime covered by de Sitter slicing (shaded in blue) in the spatial section of global AdS (cylinder).}
\label{fig:ads-ds-slicing}
\end{figure}
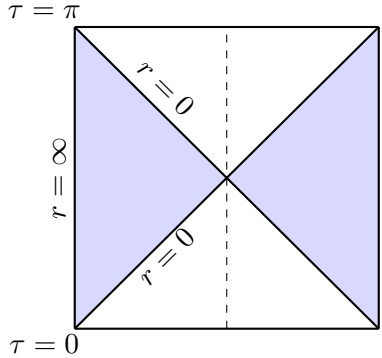

These global dS slices covers only a part global AdS spacetime as shown in figure \ref{fig:ads-ds-slicing}. Each dS slice cover half of the AdS period and infinitely many patches are required to cover the full global AdS. As $r\rightarrow0$ the warp factor vanishes i.e. the metric to become degenerate, whereby the timelike Killing vector $T\equiv\partial_{t}$ becomes null. Consequently the $r=0$ hypersurface forms an Killing horizon analogous to, though not coincident with Poincar\'e horizon.


In this parametrization of global AdS the spatial infinity is at $r\rightarrow\infty$. If we consider a bulk IR regulator, i.e. a cutoff surface at constant large $r=\Lambda$ near the AdS boundary then this hypersurface will be described by an global de Sitter spacetime. In other words, the asymptotic boundary of AdS spacetime is a global $d$-dimensional de Sitter spacetime on which the boundary CFT$_d$ lives. This foliation thus naturally leads us to the arena to study features of a CFT living in global de Sitter spacetime. In this work we employ this to study complexity of CFT living on global de Sitter, as we can simply employ the two earliest holographic complexity conjectures, namely complexity$=$volume \cite{Susskind:2014rva}  and complexity$=$action \cite{Brown:2015bva,Brown:2015lvg},which have been put forward for a field theory supported on the boundary of the holographic bulk dual spacetime. 


This set up is closely related to the dS braneworld field theory set up considered in \cite{Hawking:2000da, Iwashita:2006zj} except for one crucial distinction. In the set up of \cite{Hawking:2000da, Iwashita:2006zj}, a brane is inserted at constant $r=r_{0}$ slice, at which the bulk geometry is truncated. On the brane the intrinsic geometry is that of de Sitter spacetime. The brane has a nonzero tension and it contributes nontrivially to the entanglement entropy of field theory living on the brane. In the presence of brane the bulk theory is not only described by the Einstein-Hilbert action but it also include a brane action,
\begin{align}
    S&=S_{EH}+S_{brane}\nonumber\\
    S&=\frac{1}{16\pi G_{N}^{d+1}}\int d^{d+1}x\sqrt{-g}(R-2\Lambda)-\frac{1}{8\pi G_{N}^{d+1}}\int d^{d}x \sqrt{-h}~ T
\end{align}
In that sense, the de Sitter spacetime does not arise merely as a slicing of the bulk but instead acquires physical significance as an effective emergent geometry of the brane due to the backreaction of the the background geometry due to non-zero brane tension.

\begin{figure}
    \centering
\begin{tikzpicture}[scale=4]

\fill[blue!15] (0,1) -- (0,2) -- (0.5,1.5) -- cycle;

\draw[thick] (0,0) -- (0,2);
\draw[thick] (0,0) -- (1,1);
\draw[thick] (0,2) -- (1,1);
\draw[thick] (0,1) -- (1,1);
\draw[thick] (0,1) -- (0.5,1.5);

\node[rotate=90] at (-0.1,1.3) { $Z=0$};
\node at (-0.1,1.7) { $\uparrow$};
\node at (-0.1,1.6) { $T$};
\node[rotate=45] at (0.3,1.2) {$Z=T$};
\node at (0.75,1.5) {$\mathcal{I}^+$};
\node at (0.5,0.3) {$\mathcal{I}^-$};

\end{tikzpicture}

\caption{Penrose diagram of the Poincare patch of AdS showing the region(shaded in yellow) covered by the flat de Sitter slicing.}
\label{fig:ads-ds-flatslicing}

\end{figure}
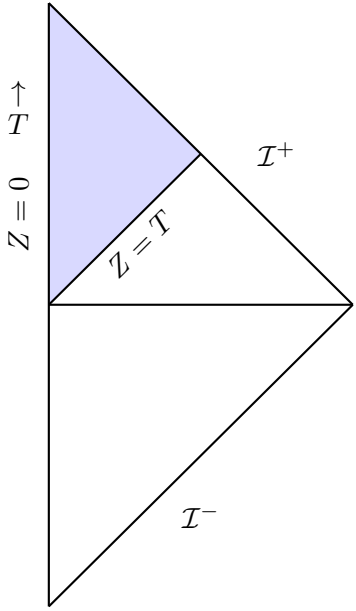

An analogous study has also been conducted in \cite{Reynolds_2017} where the authors have considered the slicing of (a part of the) Poincar\'e patch of AdS by a patch of de Sitter spacetime, namely the half of dS space as covered by conformal coordinates,
\begin{align}
    ds^{2}=\frac{1}{Z^{2}}(dZ^{2}-dT^{2}+d\vec{X}^{2})\longrightarrow ds^{2}=dr^{2}+\sinh^{2}{r} \left(\frac{-d\eta^{2}+d\vec{x}^{2}}{\eta^{2}}\right).
\end{align}
The part of Poincar\'e patch of AdS covered by these patches of dS space as foliations is displayed in figure \ref{fig:ads-ds-flatslicing}. In this set up, the flat boundary of Poincar\'e AdS spacetime is replaced by the de Sitter patches, i.e. the dual CFT lives on a patch of dS spacetime. This is in contrast to our choice of \emph{global} de Sitter slicing. This difference is quit significant as only the global de Sitter slicings enable us to probe the physics of (conformal) quantum fields in global de Sitter spacetime rather than a patch of de Sitter spacetime.

\section{Volume complexity of CFT in global de Sitter spacetime}\label{sec: Volume}
In this section we compute CFT complexity in global de Sitter spacetime by following the \emph{Complexity}=\emph{Volume} prescription, which states that the complexity (at a given time) of the $d$-dimensional boundary theory in a non-gravitational background is given by the volume of the maximal volume spacelike hypersurface ($\Sigma$) in the $(d+1)$-dimensional bulk which is anchored at the boundary,

\begin{equation}
    \mathcal{C}_{V}=\frac{V_{\Sigma}}{l\,G_{N}}
\end{equation}
To proceed with the computation of complexity, we require the induced metric on the prospective spacelike hypersurface $t(r)$,
 \begin{equation}
    ds^{2}=\left(1-t'(r)^2\sinh ^{2}r \right)\, dr^2 +\sinh ^{2}r \cosh^{2} t(r)\,d\Omega_{d-1}^{2}.
\end{equation}
The volume of this spacelike slice is then given by,
\begin{eqnarray}\label{eq: ext_volume}
    V=\int_{0}^{2\pi}d\phi\int dr (\sinh r\,\cosh t(r))^{d-1} \sqrt{1-\sinh ^2r\,t'(r)^2}
\end{eqnarray}
Since the volume functional depends on $t(r)$, there are no (obvious) first integrals. The extremization of this volume functional leads to the Euler-Lagrange equation,
\begin{eqnarray}\label{eq: euler}
     t''(r)-d \cosh(r)\sinh(r) ~t'(r)^3+(d+1) \coth(r) t'(r)-(d-1) \tanh (t(r))t'(r)^2 \nonumber\\+(d-1) \text{csch}^2(r) \tanh (t(r))=0.\label{ELG}
\end{eqnarray}
This is a nonlinear second order ordinary differential equation and has to be supplemented with two boundary conditions at the cutoff surface $r=\Lambda$, given by $t(\Lambda)=t_{\star}$ and $t'(\Lambda)=\alpha$. Furthermore, as the extremal surface must be spacelike everywhere, \emph{i.e.} the normal vector $n_{\mu}=\{t'(r),1,\vec{0}\}$ should be timelike everywhere, puts a constraint on the slope $t'(\Lambda)=\alpha$,
\begin{align}
    |\vec{n}|^{2}&<0\nonumber\\
    t'(r)^2-\text{csch}^2(\Lambda)&<0.   
\end{align}
As $\text{csch}(r)$ is a monotonically decreasing function of $r$, it is justified to choose $t'(\Lambda)=\alpha=0$ at the cutoff surface which is near the asymptotic boundary at $r\rightarrow\infty$. It can be seen from the volume functional \ref{eq: ext_volume} that $t'(r)=0$ indeed maximize the volume functional. 
Another simplification to the boundary condition can be made by choosing $t_{\star}=0$. With these simplified boundary conditions the Euler Lagrange equation accepts a trivial solution $t(r)=0\,\forall r$. Also notice that due to the specific nature of time-dependence of metric the spacetime is time-reflection symmetric only at a special time $t=0$. For this special $t(r)=0$ extremal surface we can obtain a closed form analytical expression for the maximal volume,
\begin{eqnarray}\label{eq: volume0}
    V_{\Sigma}&=&\int d\Omega_{d-1}\int_{0}^{\Lambda}dr~\sinh^{d-1}{r}\nonumber\\
    &=&\frac{2 \pi ^{d/2} \sinh ^d(\Lambda ) \, _2F_1\left(\frac{1}{2},\frac{d}{2};\frac{d+2}{2};-\sinh ^2(\Lambda )\right)}{d~\Gamma \left(\frac{d}{2}\right)}
\end{eqnarray}
When the bulk IR cutoff surface is taken to infinity \emph{i.e.} in large $\Lambda$ limit the UV-divergence structure of complexity is expected to show some universal scaling law behavior dictated by (CFT) extensivity. The bulk IR cutoff will be related to the CFT UV-cutoff, namely the CFT ``lattice spacing" via the UV-IR connection \cite{Susskind:1998dq} in AdS-CFT, $\epsilon=e^{-\Lambda}$, which is employed to translate the $\Lambda\rightarrow\infty$ limit into $\epsilon\rightarrow0$. The asymptotic behavior of the hypergeometric function involved in the complexity expression is different for even and odd values of spacetime dimension $d$ which expressed as,
\begin{align}
C_{V}&=\frac{\mathcal{V}_{d-1}}{G_{N}^{d+1}}\left(\frac{1}{(d-1) 2^{d-1}\epsilon ^{d-1}}-\frac{ (d-1) }{(d-3)2^{d-1}\epsilon ^{d-3}}+...+\frac{\Gamma\left(\frac{1}{2}\right) \Gamma \left(\frac{d}{2}\right)}{2\cos \left(\frac{\pi  d}{2}\right) \Gamma \left(\frac{d+1}{2}\right)}+\mathcal{O}(\epsilon)\right),
\end{align}
for even dimensions, and 
\begin{align}
    C_{V}&=\frac{\mathcal{V}_{d-1}}{G_{N}^{d+1}}\left(\frac{1}{(d-1) 2^{d-1}\epsilon ^{d-1}}-\frac{ (d-1) }{(d-3)2^{d-1}\epsilon ^{d-3}}+...+\frac{(\cos(d\pi))^{\frac{d+1}{2}}\Gamma\left(\frac{d}{2}\right) }{\Gamma\left(\frac{1}{2}\right)\Gamma \left(\frac{d+1}{2}\right)}\ln(\epsilon)+\mathcal{O}(\epsilon)\right)
\end{align}
for odd dimensions.

These expression of complexity exhibit the universal behavior of leading divergence of complexity as,
\begin{equation}
    C_V=  \# \, c\,\frac{\mathcal{V}_{d-1}}{\epsilon^{d-1}} +\ldots
\end{equation}
where $c$ is the central charge, $c\sim\frac{1}{G_N}$, $\mathcal{V}_{d-1}$ is the $d-1$-dimensional boundary volume (here it is the volume of the unit sphere $S_{d-1}$), and $\#$ is a spacetime dimension dependent numerical factor. The complexity is directly proportional to central charge which is a measure of the number of degrees of freedom per lattice, and the leading divergence in IR cutoff is $\epsilon^{d-1}$ which is same as $e^{(d-1)\Lambda}$. This is consistent with the fact that one is computing the complexity of \emph{local} theory, namely the dual CFT living on the dS space, and hence the complexity is expected to scale extensively with the spatial volume, which in this case is the volume of the spatial section of dS$_d$, namely the unit sphere $S_{d-1}$. Also it can be seen from these expressions that for odd dimensional de Sitter spacetime there is a logarithmic divergence with a universal numeric factor.\\

Next we turn to the more general case of non-zero $t_{\star}$ which will shed light on the dependence of complexity over the boundary time $t_{\star}$. For non-zero $t_{\star}$ we do not have the benefits of time-reflection symmetry or any nontrivial integrals of motion which prevents us from finding an an analytical solution of the Euler-Lagrange equation \eqref{ELG}. So we proceed to compute the volume of the maximal volume spacelike surface numerically. The maximal volume is determined numerically as a function of the cutoff $\Lambda$ and boundary time $t_{\star}$. The numerical plots representing the behavior of the maximal volume functional under variation of $(\Lambda, t_{\star})$ are shown in the figure below.
\begin{figure}[h]
    \centering
    \begin{minipage}{0.45\textwidth}
        \centering
        \includegraphics[height=4.5cm]{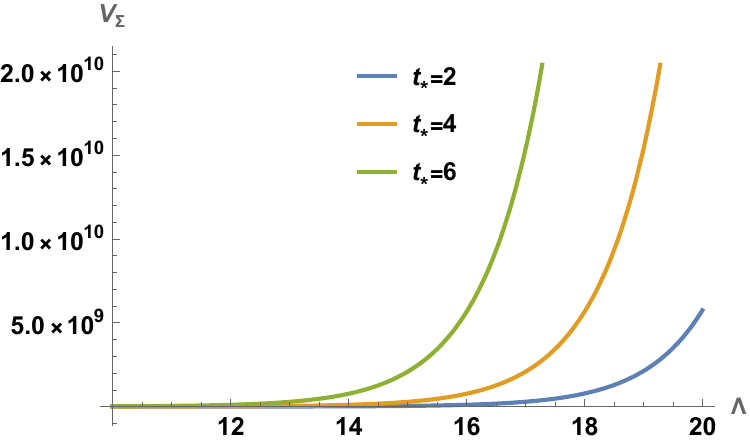}
        \caption{Maximal volume vs $\Lambda$ for $d=2$.}
        \label{fig:VvsLambda}
    \end{minipage}%
    \hfill
    \begin{minipage}{0.45\textwidth}
        \centering
        \includegraphics[height=4.5cm]{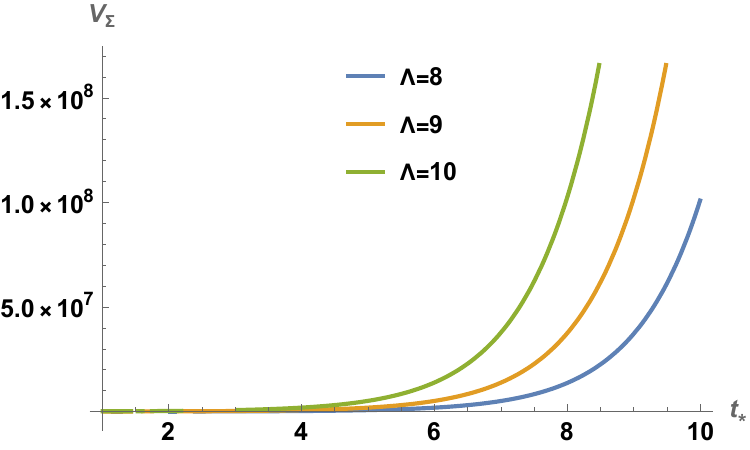}
        \caption{Maximal volume vs $t_{\star}$ for $d=2$.}
        \label{fig:Vvsts}
    \end{minipage}
    \vskip\baselineskip
    \begin{minipage}{0.45\textwidth}
        \centering
        \includegraphics[height=4.5cm]{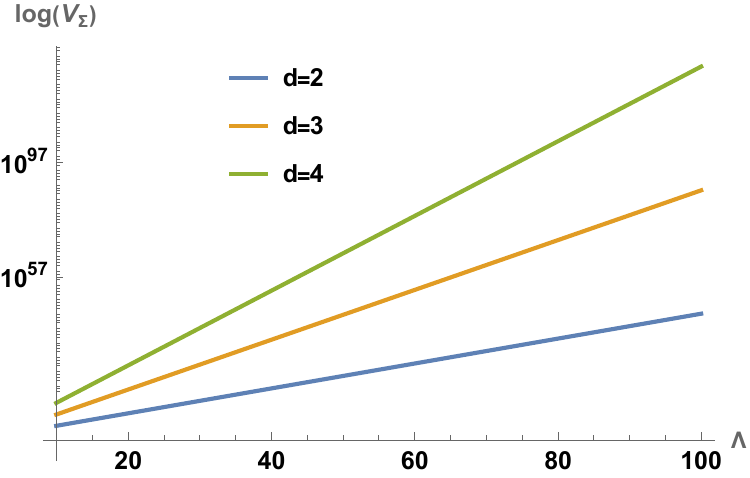}
        \caption{$\ln{V_{\Sigma}}$ vs $\Lambda$ for $d=2,3,4.$}
        \label{fig:logVvsLambda}
    \end{minipage}
    \hspace{1cm}
    \begin{minipage}{0.45\textwidth}
        \centering
        \includegraphics[height=4.5cm]{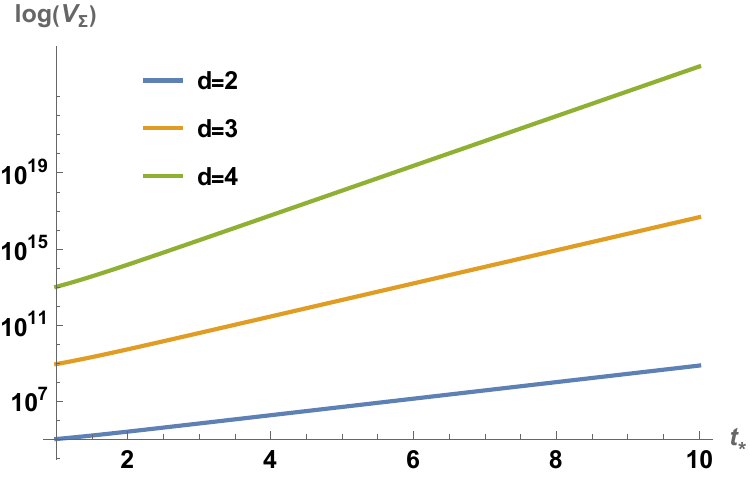}
        \caption{$\ln{V_{\Sigma}}$ vs $t_{\star}$ for $d=2,3,4.$}
        \label{fig:logVvsts}
    \end{minipage}
\end{figure}
\begin{itemize}
\item  The linear character of the semi-log plots in figure \ref{fig:logVvsLambda} and \ref{fig:logVvsts} are clear indication of exponential dependence of maximal volume on $\Lambda$ and $t_{\star}$.
    \item From the plots of figure \ref{fig:logVvsLambda} and \ref{fig:logVvsts}, we can measure the slopes (scaling exponents). They are tabulated in tables \ref{tab:logVvsLam} and \ref{tab:logVvsts} respectively.
    \begin{table}[h!]
    \centering
    \begin{minipage}{0.45\textwidth}
        \centering
        \begin{tabular}{|c|c|}
    \hline
 Dimension ($d$)& slope=$\frac{\Delta\ln(V_{\Sigma})}{\Delta \Lambda}$\\
    \hline
      2  & 1.00003  \\
      \hline
      3  & 2.00001  \\
      \hline
      4  & 3.00000\\
      \hline
    \end{tabular}
    \caption{Index of $\Lambda$ dependence in $V_{\Sigma}$ }
    \label{tab:logVvsLam}
    \end{minipage}%
    \hspace{0.05\textwidth} 
    \begin{minipage}{0.45\textwidth}
        \centering
        \begin{tabular}{|c|c|}
    \hline
 Dimension ($d$)& slope=$\frac{\Delta\ln(V_{\Sigma})}{\Delta t_{\star}}$\\
    \hline
      2  & 0.99393  \\
      \hline
      3  & 1.98795  \\
      \hline
      4  & 2.98193\\
      \hline
    \end{tabular}
    \caption{Index of $t_{\star}$ dependence in $V_{\Sigma}$ }
    \label{tab:logVvsts}
    \end{minipage}
\end{table}

\item The slopes indicates a dependence $V_{\Sigma}\sim e^{(d-1)\Lambda}e^{(d-1)t_{\star}}$. This agrees with our expectation that complexity to be proportional to the spatial volume. As we have $d$-dimensional de Sitter spacetime whose spatial volume, in units of the cutoff, is proportionals to $V_{\Sigma}\sim \sinh^{d-1}{\Lambda}\cos^{d-1}{t_{\star}}$.  It is worth noticing that the complexity grows exponentially with time. This exponential growth can be understood as an artifact of the inflationary growth of spatial sections of global de Sitter spacetime. In general, the expectation is that complexity grows due to an increment in the quantum entanglement between the degrees of freedom of the system but this is not the case here. Here, as the spatial volume of the de Sitter spacetime grows exponentially with time (due to presence of inflationary scale factor $\cosh{t}$) the number of degrees of freedom involved also increase (a Hubble volume worth of degrees of freedom added with every e-folding of inflation). As a result, the complexity of a local theory on this inflating background, exhibits an exponential growth with time (since the complexity of a local CFT living on a spacetime tracks the spatial volume). This exponential growth of complexity indicates contrast with the Lloyd bound on quantum computational complexity \cite{Lloyd:2000cry}. This might be an indication towards violation of the Llyod bound for systems where degree of freedom are being added during time-evolution.

\item Another point to note here is that the volume complexity does not exhibit \textit{hyperfast} growth i.e. it does not diverge at finite time, which was previously observed for the static patch dS holography \cite{Jorstad:2022mls}. In this regard, we note that the holographic complexity that we study are of different CFT's living on a different spacetime. In \cite{Jorstad:2022mls} they study complexity of a CFT that lives on the stretched horizon of the de Sitter static patch, which is conjectured dual to the dynamical quantum gravity theory in de Sitter spacetime. In contrast we have found the holographic complexity of a CFT which lives on a undeformed or unbackreacted global de Sitter spacetime i.e. without including dS gravitational degrees of freedom.

\item  Finally we compare our findings with that in \cite{Reynolds_2017} where the complexity of a CFT living on a patch (half) of de Sitter spacetime was computed. They have found similar results where the complexity is time-dependent and is proportional to the spatial volume of the boundary\emph{i.e.}
\begin{align}
    C\sim V_{x}\left(\frac{\sinh\rho_{0}}{\eta_{0}}\right)^{d-1}
\end{align} 
where, $\eta=\eta_{0}$ is boundary time in conformal de Sitter and $\rho_{0}$ is the boundary IR cutoff.  Using the map between the conformal and global de Sitter foliation coordinates given explicitly in appendix \ref{sec: ads_ds_coord2} it can be seen that the dependence of complexity over IR cutoff is in agreement with our results,
\begin{align}
   C\sim \left(\sinh{\Lambda} \right)^{d-1}
\end{align}
however since the constant conformal time slice $\eta=\eta_{0}$ does not coincide with the constant global time $t$ surface, a direct diffeomorphism between the foliations is not expected to exhibit an agreement in the complexity. It is important to notice that although both the global de Sitter geometry considered in this work and conformal de Sitter geometry considered in \cite{Reynolds_2017} span over a part of global AdS spacetime, the vacuum of the CFT living over conformal de Sitter and global de Sitter spacetime are different, consequently the states over which the complexity are measured are not equivalent, and the complexity result are not expected to be matched.

\end{itemize}

\section{Action complexity of CFT in global de Sitter spacetime}\label{sec: action}
In this section we compute the complexity of CFT living on de Sitter spacetime, following the \textit{complexity}= \textit{action} proposal, which identifies the complexity of the boundary field theory with the bulk onshell action over Wheeler deWitt (WDW) patch, 
\begin{equation}
    C_{A}=\frac{I_{WDW}}{\pi \hbar}.
\end{equation}
Wheeler deWitt patch is the union of all the spacelike hypersurfaces anchored at a boundary time say $t_{\star}$ bounded by null hypersurfaces, in other words it is the domain of dependence of the time slice that has been considered in the complexity= volume proposal. Here, the bulk $(d+1)$-dimensional gravitational theory is described by the Einstein-Hilbert action with a negative cosmological constant $\Lambda$,
\begin{eqnarray}
    S=\frac{1}{16\pi G^{d+1}_{N}}\int d^{d+1}x \sqrt{-g} \left(R-2\Lambda\right)
\end{eqnarray}
which admits the maximally symmmetric pure AdS$_{d+1}$ spacetime as an solution. In global de Sitter foliations,
\begin{equation}
    ds^{2}=dr^{2}+\sinh^{2}{r}\,(-dt^{2}+\cosh^{2}{t}~d\Omega_{d-1}^{2}).
\end{equation}
One should bear in mind that these de Sitter foliations does not cover the full AdS spacetime, it only cover a part of global of AdS spacetime, so the support of the WDW patch action integral is only confined to that part of global AdS. To regulate the action, the WDW patch is fixed at the boundary cutoff $r=\Lambda$ at the boundary time $t=t_{\star}$. Then the null curves which constitute the null boundaries of the WDW patch are given by,
\begin{eqnarray}
    ds^{2}&=&0\nonumber\\
    \pm \int_{t_{\star}}^{t} dt&=&\int_{\Lambda}^{r}\frac{dr}{\sinh{r}}\nonumber\\
t&=&t_{\star}\pm\left(\ln \left(\coth \left(\frac{r }{2}\right)\right)-\ln \left(\coth \left(\frac{\Lambda}{2}\right)\right)\right).
\end{eqnarray}\label{eq:null_hypsurf}
Although these null trajectories may appear different from the null trajectories in the global AdS coordinates, they are related by diffeomorphism, the only distinction arises due to the termination of these null trajectories at the null horizon, where the de Sitter foliations cease to exist.
Having determined the boundaries of the WdW patch, we proceed to computing the onshell action over the WDW patch, namely
\begin{eqnarray}
    I_{WdW}=I_{bulk}+I_{bdry}+I_{joint}.
\end{eqnarray}
The bulk contribution comes from the Einstein-Hilbert action,\footnote{To reduce clutter of notation, for the action complexity section, we adopt the convention $16\pi G_{N}^{d+1} S=I$},
\begin{eqnarray}
    I_{EH}&=&\int d^{3}x \sqrt{-g}(R-2\Lambda)
\end{eqnarray}
The Ricci Scalar for $\text{AdS}_{d+1}$ is given by,
\begin{equation}
    R=-d(d+1)
\end{equation}
with cosmological constant,
\begin{equation}
    \Lambda_{AdS}=-\frac{d(d-1)}{2}
\end{equation}
\textbf{Einstein Hilbert part of action:}
\begin{eqnarray}
    I_{EH}&=&\int_{0}^{\Lambda}dr\int_{t_{\star}-t(r)}^{t_{\star}+t(r)} dt\int d\Omega_{d-1}\sqrt{-g}(R-2\Lambda_{AdS})\nonumber\\
    &=&-\frac{4 d \pi ^{d/2} }{\Gamma \left(\frac{d}{2}\right)}\int_{0}^{\Lambda}dr~\sinh ^d(r)\int dt \cosh ^{d-1}(t)
\end{eqnarray}
Instead of working for general $d$, it turns out to be more convenient to separate into two subcases, namely odd and even $d$ since the series expansions are rather different (e.g. one includes a log divergence while the other doesn't). For an odd dimensional de Sitter boundary spacetime \emph{i.e.} when $d$ is odd,
\begin{eqnarray}
    I_{EH}=&=&-\frac{4 d \pi ^{d/2} }{\Gamma \left(\frac{d}{2}\right)}\int_{0}^{\Lambda}dr~\sinh ^d(r)\int dt \cosh ^{d-1}(t)\nonumber\\
    &=&-\frac{4 d \pi ^{d/2} }{\Gamma \left(\frac{d}{2}\right)}\Bigg(C^{\frac{d-1}{2}}_{d-1}\frac{2}{2^{d-1}}I_{1}+\frac{1}{2^{d-2}}\sum_{k=0}^{\frac{(d-3)}{2}}C^{k}_{d-1}\frac{1}{\left(d-1-2k\right)}\cosh[(d-2k-1)t_{\star}]\nonumber\\&&\times\Bigg(\left(\coth\left(\frac{\Lambda}{2}\right)\right)^{-(d-2k-1)}I_{2}-\left(\coth\left(\frac{\Lambda}{2}\right)\right)^{(d-2k-1)}I_{3}\Bigg).
\end{eqnarray}
For an even dimensional de Sitter spacetime \emph{i.e.} when $d$ is even number,
\begin{eqnarray}
    I_{EH}    
    &=&-\frac{4 d \pi ^{d/2} }{\Gamma \left(\frac{d}{2}\right)}\sum_{k=0}^{\frac{d}{2}-1}C^{k}_{d-1}\frac{1}{2^{d-2}(d-2k-1)}\cosh[(d-2k-1)t_{\star}]\nonumber\\&&\times\Bigg(\left(\coth\left(\frac{\Lambda}{2}\right)\right)^{-(d-2k-1)}I_{2}-\left(\coth\left(\frac{\Lambda}{2}\right)\right)^{(d-2k-1)}I_{3}\Bigg),
\end{eqnarray}
where, $I_{1}$, $I_{2}$ and $I_{3}$ are some integral forms given by,
\begin{align}
I_{1}=&\int_{0}^{\Lambda} dr~\sinh^{d}(r)\ln\left(\frac{\coth\left(\frac{r}{2}\right)}{\coth\left(\frac{\Lambda}{2}\right)}\right)\nonumber\\
I_{2}=&\int_{0}^{\Lambda} dr~(\sinh r)^{2d-2k-1}(\cosh r-1)^{-(d-2k-1)}\nonumber\\
    I_{3}=&\int_{0}^{\Lambda} dr~(\sinh r)^{(2k+1)}(\cosh r-1)^{(d-2k-1)}\nonumber
\end{align}
The explicit result of these integrals have been shown in appendix \ref{app:integrals}. Also the total contribution of these integral results can be combined into $I_{EH}$, but the result for general $d$ is quit long therefore we refrain from presenting the gory details here. But we can present the leading order dependence of these terms on the IR cutoff $\Lambda$ which goes as,
\begin{align}
   I_{EH}^{\text{odd}}&\sim-\frac{4 d \pi ^{d/2} }{\Gamma \left(\frac{d}{2}\right)}\Bigg(C^{\frac{d-1}{2}}_{d-1}\frac{2}{2^{d-1}}\left(\frac{ 2^{1-d} e ^{(d-1)\Lambda}}{(d-1) d}-\frac{(-1)^{\frac{d-1}{2}} \Gamma \left(\frac{3}{2}\right) \Gamma \left(\frac{d+1}{2}\right)}{\Gamma \left(\frac{d+2}{2}\right)}\Lambda\right)\nonumber\\&+\frac{1}{2^{d-2}}\sum_{k=0}^{\frac{(d-3)}{2}}C^{k}_{d-1}\frac{1}{\left(d-2k-1\right)}\cosh[(d-2k-1)t_{\star}]\frac{2^{2-d} (d-2 k-1) e ^{(d-1)\Lambda}}{(d-1) d}\Bigg), \\
       I_{EH}^{\text{even}}&\sim-\frac{4 d \pi ^{d/2} }{\Gamma \left(\frac{d}{2}\right)}\sum_{k=0}^{\frac{d}{2}-1}C^{k}_{d-1}\frac{1}{2^{d-2}(d-2k-1)}\cosh[(d-2k-1)t_{\star}]\frac{2^{2-d} (d-2 k-1) e ^{(d-1)\Lambda}}{(d-1) d}.
\end{align}
From this asymptotic behavior it can be noticed that the leading contribution from the Einstein Hilbert action is propositional to the spatial volume of the de Sitter slice which goes as $e^{(d-1)\Lambda}$ near the asymptotic boundary. Since the full action complexity is determined by other contributions, there is no point in dwelling on the Einstein-Hilbert contribution in isolation. For the 2-dimensional de Sitter spacetime ($d=2$) the the EH part of the action is given by,
\begin{eqnarray}
    I_{EH}&=&\int_{0}^{2\pi} d\phi\int_{0}^{\Lambda} dr\int_{-t(r)}^{t(r)}\sqrt{-g}(R+2)\nonumber\\
    &=&-16 \pi \cosh (t_{\star}) \sinh ^2\left(\frac{\Lambda }{2}\right) \tanh \left(\frac{\Lambda }{2}\right) 
\end{eqnarray}
which for the special case of $t_{\star}=0$ is,
\begin{align}
    I_{EH}
    &=-16 \pi \sinh ^2\left(\frac{\Lambda }{2}\right) \tanh \left(\frac{\Lambda }{2}\right) 
\end{align}
\textbf{Boundary GHY contributions:}
\newline
\\
There will be two boundary GHY term contributions to the boundary action over WdW patch coming from the future and past null hypersurfaces denoted by $\mathcal{N_{+}}$ and $\mathcal{N}_{-} $,
\begin{eqnarray}
  I_{bdry}&=&I_{\mathcal{N}_{+}}+I_{\mathcal{N}_{-}}\nonumber\\ 
  I_{\mathcal{N}_{\pm}}&=&-(\pm)2\int d\phi d\lambda~\kappa
\end{eqnarray}
where, $\lambda$ is a parameter on null generators of $\mathcal{N}_{\pm}$ and 
\begin{align}
 \kappa n^{\mu}=n^{\nu}\nabla_{\nu}n^{\mu} 
\end{align} 
where $n^{\mu}$ is the tangent vector $n^{\mu}=\frac{\partial x^{\mu}}{\partial\lambda}$ on the null hypersurface. To simplify the calculation we can choose the null parameter $\lambda$ so that $\kappa$ vanishes, which will lead to a vanishing contribution to the boundary action. To choose such $\lambda$, let us consider a vector at the null hypersurface $x^{\mu}=\{r,\pm t(r),\phi\}$, then the tangent vector is given by 
\begin{align}
  n^{\mu}= \left(\frac{1}{\lambda'(r)},\pm \frac{t'(r)}{\lambda'(r)},0 \right). 
\end{align} 
Then solve for $\lambda$ so that $\kappa=0$ \emph{i.e.},
\begin{equation}
    n^{\nu}\nabla_{\nu}n^{\mu}=0,\nonumber
\end{equation}
this gives a single equation for $\lambda(r)$
\begin{equation}
    \coth (r) \lambda '(r)-\lambda ''(r)=0
\end{equation}
for the null hypersurface $t(r)=t_{\star}\pm(\ln(\coth{\frac{r}{2}})-\ln(\coth{\frac{\Lambda}{2}}))$, which has a solution,
\begin{align}
 \lambda(r)=c_{1}+c_{2}\cosh{r}   
\end{align}
Therefore any choice of that is of form $\lambda(r)=c_{1}+c_{2}\cosh{r}$ gives $\kappa=0$, leading to a vanishing boundary contribution to the WDW action. We make the choice $\lambda(r)=-\cosh{r}$ for future null surface $\mathcal{N}_{+}$ and $\lambda(r)=\cosh{r}$ for past null surface $\mathcal{N}_{-}$. Apart from these vanishing GHY term contributions, there are other nonvanishing WDW boundary contributions to the action, which are the so called LMPS terms which are worked out in the following.
\newline
\\
\textbf{Joint or Edge contributions:}
\newline
\\
The joint contribution to the action comes from the intersection of two null hypersurface at the cutoff surface $r=\Lambda$,
\begin{eqnarray}
    I_{\text{joint}}=2\text{sign}_{m}\int d^{d-2}x \sqrt{\gamma}~ a_{m}
\end{eqnarray}
where sign$_{m}=1$ if the volume of interest lie in future of the joint point $m$ and $m$ lies past of the null boundary, otherwise sign$_{m}=-1$, $a$ is given by
\begin{align}
    a=\ln\left({\frac{k.\bar{k}}{2}}\right)
\end{align}
and $k$ and $\bar{k}$ are the normal vectors on the null hypersurface $\mathcal{N}_{\pm}$. Also $\gamma$ is the induced metric at the joint surface which for our case is given by,
\begin{eqnarray}
    \gamma_{\mu\nu}&=&\sinh^{2}{\Lambda}\cosh^{2}{t_{\star}}~d\Omega_{d-1}^{2}.
\end{eqnarray}
The joint contribution is given by,
\begin{align}
    I_{\text{joint}}=&-2\int d\Omega_{d-1}\sqrt{\gamma}\ln\left(-\frac{k.\bar{k}}{2}\right)\nonumber\\
   =&-2\int d\Omega_{d-1}(\sinh{\Lambda}\cosh{t_{\star}})^{d-1}~\ln\left(\text{cosech}^{2}{\Lambda}\right)\nonumber\\
    =&\frac{8\pi^{\frac{d}{2}}}{\Gamma\left(\frac{d}{2}\right)}(\sinh{\Lambda}\cosh{t_{\star}})^{d-1}\ln\left(\sinh{\Lambda}\right)\nonumber\\
\end{align}
For the special case of two dimensional de Sitter spacetime with $t_{\star}=0$ the joint contribution simplifies to,
\begin{eqnarray}
    I_{joint}&=&-2\int_{0}^{2\pi}d\phi\ln\left(-\frac{k.\bar{k}}{2}\right)\nonumber\\
    I_{joint}&=&8\pi(\sinh{\Lambda}) \ln(\sinh{\Lambda})
\end{eqnarray}\\
\textbf{Boundary LMPS term contribution:}
\newline
\\
In addition to the boundary and joint contributions to the action over the WDW patch we also need to add an counter term or the LMPS term  to avoid the ambiguity associated with reparameterization of $\lambda$ \cite{Lehner:2016vdi}. This counter term is given by,
\begin{eqnarray}
    I_{LMPS}=-2\text{sign}(N_{i})\int d\phi \sqrt{\gamma} d\lambda~ \Theta\ln(|\Theta|)
\end{eqnarray}
where $\Theta$ is the expansion parameter, 
\begin{align}
    \Theta=\frac{1}{\sqrt{-\gamma}}\frac{\partial\sqrt{-\gamma}}{\partial \lambda}
\end{align} 
and the $\text{sign}(N_{i})=+1$ if the null surface lies to the future of volume of interest and $\text{sign}(N_{i})=-1$ if the null surface lies to the past of volume of interest.
\begin{eqnarray}
    I_{LMPS}
    &=&-2\int d\Omega_{d-1} \int d\lambda_{F}~ \sqrt{\gamma_{F}}\Theta_{F}\ln(|\Theta_{F}|)+2\int d\Omega_{d-1}\int d\lambda_{P}~ \sqrt{\gamma_{P}}\Theta_{P}\ln(|\Theta_{P}|)\nonumber
\end{eqnarray}
As we have discussed previously that we fix the parameter $\lambda$ as
\begin{align}
    \lambda_{F}&=-\cosh{r},\quad\lambda_{P}=\cosh{r}\nonumber
\end{align}
which gives,
\begin{align}
    \Theta_{F}=&\frac{(d-1) (\cosh (\Lambda ) \cosh t_{\star}-\sinh t_{\star})}{\cosh t_{\star} (\lambda_{F}  \cosh (\Lambda )+1)-\sinh t_{\star} (\lambda_{F} +\cosh (\Lambda ))},\nonumber\\
    \Theta_{P}=&\frac{(d-1) (\cosh (\Lambda ) \cosh (t_{\star})+\sinh (t_{\star}))}{\cosh (t_{\star}) (\lambda_{P}  \cosh (\Lambda )-1)+\sinh (t_{\star}) (\lambda_{P} -\cosh (\Lambda ))}.
\end{align}
Therefore the LMPS term is given by,
\begin{align}
    I_{LMPS}&=-2\int d\Omega_{d-1} \int d\lambda_{F}~ \sqrt{h_{F}}\Theta_{F}\ln(|\Theta_{F}|)+2\int d\Omega_{d-1} \int d\lambda_{P}~ \sqrt{h_{P}}\Theta_{P}\ln(|\Theta_{P}|)\nonumber\\
    &=\frac{4\pi^{\frac{d}{2}}}{\Gamma\left(\frac{d}{2}\right)}\tanh^{d-1}\left(\frac{\Lambda}{2}\right)\Bigg(\sinh ((d-1) t_{\star}) \ln \left(\frac{e^{2 t_{\star}} (\cosh (\Lambda )+\tanh (t_{\star}))}{\cosh (\Lambda ) -\tanh (t_{\star})}\right)\nonumber\\& -\cosh((d-1)t_{\star})\left(\frac{2}{d-1}-2\ln\left(\frac{\cosh{\Lambda}-1}{d-1}\right)+\ln(\left(\cosh{^{2}{t_{\star}}\cosh^{2}{\Lambda}}-\sinh^{2}{t_{\star}}\right)\right)\nonumber\\&+\cosh^{d-1}{t_{\star}}(\cosh{\Lambda}+1)^{d-1}\Big(\frac{2}{d-1}-4\ln(\sinh{\Lambda})\nonumber\\&+\ln((d-1)^{2}(\cosh^{2}{\Lambda}-\tanh^{2}{t_{\star}}))\Big)\Bigg)
\end{align}
At this point we should emphasize that for the special case of $t_{\star}=0$, this contribution of LMPS term simplifies drastically. This is due to the fact that the bulk metric or the global de Sitter metric,
\begin{align}
    ds^{2}=dr^{2}+\sinh^{2}{r}(-dt^{2}+\cosh^{2}{t}d\Omega_{d-1}^{2})
\end{align}
despite depending explicitly on time $t$ and therefore not having an time translation invariance, still has a time reflection symmetry across $t=0$ therefore when the WDW patch is anchored at $t_{\star}=0$ the contributions to the action also respect this symmetry. This is clear from the expressions of $\Theta$'s, as they satisfy $\Theta_{F}=-\Theta_{P}$ for $t_{\star}=0$. This simplification leads to the identical contribution to the LMPS term from future and past null boundaries.
\begin{eqnarray}
 I_{LMPS} &=&-4\int d\Omega_{d-1} \int d\lambda_{F}~ \sqrt{h_{F}}\Theta_{F}\ln(|\Theta_{F}|)\nonumber
\end{eqnarray}
For the two dimensional de Sitter spacetime this simplifies to,
\begin{align}
    I_{LMPS} =&8 \pi  \tanh \left(\frac{\Lambda }{2}\right) (\cosh (\Lambda )+\ln (1-\text{sech}(\Lambda ))-2(\cosh (\Lambda )+1) \ln (\sinh (\Lambda ))\nonumber\\&+(\cosh (\Lambda )+1) \ln (\cosh (\Lambda )))\nonumber
\end{align}

Another rather technical point to notice is that the part of second last terms(~$\ln(\sinh(\Lambda))$ in the LMPS term cancels out the contribution from the joint term. Therefore simplifies the combined contribution to the WDW action to,
\begin{align}
    I_{\text{joint}}+I_{\text{LMPS}}=&=\frac{4\pi^{\frac{d}{2}}}{\Gamma\left(\frac{d}{2}\right)}\tanh^{d-1}\left(\frac{\Lambda}{2}\right)\Bigg(\sinh ((d-1) t_{\star}) \ln \left(\frac{e^{2 t_{\star}} (\cosh (\Lambda )+\tanh (t_{\star}))}{\cosh (\Lambda ) -\tanh (t_{\star})}\right)\nonumber\\& -\cosh((d-1)t_{\star})\left(\frac{2}{d-1}-2\ln\Big(\frac{\cosh{\Lambda}-1}{d-1}\right)\nonumber\\&+\ln\left(\cosh{^{2}{t_{\star}}\cosh^{2}{\Lambda}}-\sinh^{2}{t_{\star}}\right)\Big)+\cosh^{d-1}{t_{\star}}(\cosh{\Lambda}+1)^{d-1}\nonumber\\&\Big(\frac{2}{d-1}-2\ln(\sinh{\Lambda})+\ln((d-1)^{2}(\cosh^{2}{\Lambda}-\tanh^{2}{t_{\star}}))\Big)\Bigg)
\end{align}
Now we combine all the contribution to the WDW action and the result for generic $d$ dimensions can be obtained. We only present here the simple result of two dimensional de Sitter spacetime with the anchoring boundary time $t_{\star}=0$,
\begin{eqnarray}
    I_{WdW}&=&I_{EH}+I_{bdry}+I_{joint}+I_{LMPS}\nonumber\\
    &=&8 \pi  \tanh \left(\frac{\Lambda }{2}\right) \Big(1+ \cosh (\Lambda ) \ln (\cosh (\Lambda ))+ \ln (\cosh (\Lambda )-1)\nonumber\\&&- (\cosh (\Lambda )+1) \ln (\sinh (\Lambda ))\Big)
\end{eqnarray}
But in the limiting case when we take the IR cutoff $\Lambda$ to be large and again introduce $\epsilon=e^{-\Lambda}$ then the expressions get simplified and we can see the divergent structure of WDW action.
\begin{align}
    d=2,\quad I_{WDW}=&8 \pi(  t_{*} \sinh (t_{*})+  \cosh (t_{*})-  \cosh (t_{*}) \log (\cosh (t_{*})))+\mathcal{O}(\epsilon)\nonumber\\
  d=3,\quad  I_{WDW}=&\frac{4 \pi  \log (2) \cosh ^2(t_{*})}{\epsilon ^2}-16\pi \ln(\epsilon)+4 \pi  (4 t_{*} \sinh (2 t_{*})\nonumber\\&+\cosh (2 t_{*}) (2-5 \log (2)-4 \log (\cosh (t_{*})))+2-9 \log (2))+\mathcal{O}(\epsilon)\nonumber\\
 d=4,\quad I_{WDW}=&\frac{\pi ^2 \log (3) \cosh ^3(t_{*})}{\epsilon ^3}-\frac{3 \pi ^2 \cosh (t_{*}) (\log (3) \cosh^{2} ( t_{*})-2)}{ \epsilon }\nonumber\\&-\frac{8}{3} \pi ^2 (6 \cosh (t_{*})-3 t_{*} \sinh (3 t_{*})-\cosh (3 t_{*}) (1-3 \log (\cosh (t_{*}))\nonumber\\&-3\log (3)))+\mathcal{O}(\epsilon)\nonumber\\
d=5,\quad I_{WDW}=&\frac{4 \pi ^2 \log (2) \cosh ^4(t_{*})}{3 \epsilon ^4}-\frac{8 \pi ^2 \cosh ^2(t_{*}) \left(\log (4) \cosh ^2(t_{*})-1\right)}{3 \epsilon ^2}+\frac{32}{3}\pi^{2}\ln{\epsilon}\nonumber\\&-\frac{4}{3} \pi ^2 (4-8 t_{*} \sinh (4 t_{*})+10 \cosh (2 t_{*})+\log (2) (61 \cosh (2 t_{*})-67) \cosh ^2(t_{*})\nonumber\\&+2 \cosh (4 t_{*}) (4 \log (\cosh (t_{*}))-1))+\mathcal{O}(\epsilon)
\end{align}
From the diverging structure of WDW action it can be seen explicitly that the leading divergence of action complexity goes as $\epsilon^{-(d-1)}\cosh^{d-1}{t_{\star}}$ which in terms of $\Lambda$ is $e^{(d-1)\Lambda}(\cosh{t_{\star}})^{d-1}$. This is in agreement to our expectation as complexity, being an extensive quantity, is proportional to the spatial volume for a local field theory (recall that for the de Sitter spacetime the spatial volume is $\sim (\sinh{\Lambda}\cosh{t_{\star}})^{d-1}$). This leading divergence structure is also in perfect agreement from the numerical result that we obtained following complexity equals volume conjecture. 

Also, we note here in passing that for the special case of $t_{\star}=0$ an explicit comparison between the diverging structure of complexity can be obtained analytically from following volume and action conjecture. As per expectations, they display an agreement upto an overall numerical factor (
Similar agreement between the volume and action conjectures was observed for the holographic complexity of CFT living on the boundary of AdS \cite{Reynolds:2016rvl}. Furthermore notice that for an CFT living over odd dimensional de Sitter spacetime the complexity has an logarithmic divergence which is absent in case of even dimensional de Sitter spacetime. This feature has also been observed for the holographic complexity of AdS spacetime. We believe the fact that quantum complexity of a CFT living over de Sitter spacetime inherit these feature is an non-trivial result and indicates that complexity of a CFT living on de Sitter spacetime share some of the universal features of CFT living on AdS boundary. 

\section{Complexity of CFT on a dS braneworld}\label{sec: brane}

In this section we focus on a case similar to \cite{Hawking:2000da, Iwashita:2006zj} whereby we insert a brane with nonvanishing tension $T$ near the asymptotic boundary of AdS (near $r\rightarrow\infty$) \emph{i.e} the IR cutoff surface is replaced by an brane. The total bulk action describing the $(d+1)$ dimensional gravity theory is given by,
\begin{align}
    S=&S_{\text{EH}}+S_{\text{brane}}\nonumber\\
    =&\frac{1}{16\pi G}\int d^{d+1}x\sqrt{-g}(R-2\Lambda)-\frac{1}{8\pi G}\int d^{d}x\sqrt{-h}~T
\end{align}
where, $T$ is the brane tension. In this scenario, the boundary theory living on asymptotic boundary of AdS actually lives on this brane (with nonzero tension) and the extrinsic geometry on the brane in the chosen foliation of AdS spacetime is described by global de Sitter. The insertion of the brane with nonzero tension at the de Sitter slice $r=r_{B}$  give rise to dynamical theory of gravity on the brane due to the backreaction of brane tension over the de Sitter geometry. This setup is widely known as \textit{braneworld holography}\cite{deHaro:2000wj,Karch:2000ct,Randall:1999vf} as the boundary theory living on AdS boundary now resides on this brane and an effective higher derivative theory of gravity is induced on the brane \cite{Bueno:2022log} with an effective Newton's constant,
 \begin{align}\label{eq:eff_newt_const}
     G_{d}=\frac{d-2}{2}\coth{r_{B}}~G_{d+1}.
 \end{align}
 If the brane is embedded at some fixed radial position $r=r_{B}$, than the brane tension is determined by the Israel junction condition at the brane. If we glue two copy of AdS spacetime across the brane\footnote{This gluing is done for an two sided brane.} than the outward normal vectors at the brane are equal and opposite to each other. This is valid when the two copies are mirror image of each other \emph{i.e.} there is $\mathbb{Z}_{2}$ symmetry across the brane,
\begin{eqnarray}
n_{\mu}^{+}&=&-n_{\mu}^{-}\nonumber\\
K_{\mu\nu}^{\pm}&=&\pm(\coth{r})h_{\mu\nu}\nonumber\\
K^{\pm}&=&\pm 2\coth{r}.
\end{eqnarray}
  Using these expressions for the extrinsic curvature, the brane tension determined by the Israel Junction conditions,
  \begin{equation}
   ( K_{\mu\nu}-K h_{\mu\nu})|_{r>r_{B}}-( K_{\mu\nu}-K h_{\mu\nu})|_{r<r_{B}}=8\pi G_{d+1}T_{\mu\nu}.
\end{equation}
and is given by,
  \begin{eqnarray}
      T=\frac{d-1}{4\pi G_{d+1}}\coth{r_{B}}
  \end{eqnarray}
  When we place the brane near asymptotic boundary of AdS, $r_{B}\rightarrow\infty$,
  \begin{eqnarray}
           T=\frac{d-1}{4\pi G_{d+1}}
  \end{eqnarray}
  Our goal is to study effect of this dS brane insertion over the quantum complexity of CFT living on global de Sitter following both volume and action conjectures of holographic complexity. While here we focus only on the quantum complexity similar studies in presence of de Sitter brane have been conducted before in a more general context \cite{Nojiri:2002wn,Calcagni:2005vn,Emparan:2022ijy,Panella:2023lsi}. In this work, we do not consider a dynamical brane \emph{i.e} we do not consider the Einstein Hilbert gravitational action of brane into account and leave it as an future goal. In similar setup while the geometry on brane is static de Sitter, the entanglement entropy is studied in \cite{Hawking:2000da,Iwashita:2006zj}. The entanglement entropy is given by,
  \begin{eqnarray}\label{eq:Sent_brane}
      S_{\text{ent}}&=&\frac{2\Omega_{d-2}\int_{0}^{r_{B}}dr \sin^{d-2}{r}}{4G_{d+1}}
  \end{eqnarray}
  with the extremal surface $\rho=\sinh{r_{B}}$, where $\rho$ is the radial coordinate of the static de Sitter. From this expression of entanglement entropy it can be noticed that the entanglement entropy is just twice the entanglement entropy of one copy with boundary at $r_{B}$, this may happen as the entanglement entropy measures the area of extremal surface which do not get directly effected by the additional Nambu-Goto action of the brane however an indirect effect of backreaction of brane tension on the geometry might be captured by the entanglement entropy. The major finding of this de Sitter braneworld is that the entanglement entropy\footnote{Similar result for AdS brane has been found previously in \cite{Emparan:2006ni,Myers:2013lva}} is same as the entropy associated with cosmological horizon when $r_{B}\gg1$ \emph{i.e} when the brane is near boundary. Also, notably in \cite{Iwashita:2006zj} a precise agreement between this entanglement entropy and the entropy calculated from euclidean path integral has been obtained. Furthermore an study of entanglement entropy and volume complexity in the context of wedge holography has been done in \cite{Aguilar-Gutierrez:2023tic} in which two static de Sitter branes are inserted in the $(d+1)$ dimensional AdS, one near the asymptotic boundary and one inside bulk spacetime.

 Another feature of the setup to be noticed is if the brane is at some fixed non-zero value of $r=r_{B}$. Then the brane intersects the asymptotic (global) AdS boundary $\rho\rightarrow\infty$, when $\tau=0$ \emph{i.e.} $t\rightarrow\infty$ in de Sitter time. This feature is quite different from what happens in the case of AdS brane \cite{Chen:2020uac}, where the brane intersects the AdS boundary $\mathbb{R}\times S^{d-1}$ at some $\theta=\frac{\pi}{2}$ and covers half of the boundary hemisphere over which the spacetime is glued, the details of this is given in appendix \ref{app: AdS_AdS}. In our setup the de Sitter brane at fixed $r=r_{B}$ reside in global AdS spacetime at a curve $\mathcal{B}(\rho,\tau)$ satisfying, 
\begin{align}
    \sin{\tau}=\frac{\cosh{r_{B}}}{\cosh{\rho}}.
\end{align}
If we push the brane to $r_{B}\rightarrow\infty$, then in the global AdS spacetime the brane covers the full asymptotic AdS boundary rather than a part of it. Therefore it is interesting to see how this crucial difference between AdS$_{d}$ foliation of AdS$_{d+1}$ and dS$_{d}$ foliation of AdS$_{d+1}$ effects the complexity of the boundary CFT living on the brane.
  \subsection{Volume Complexity in presence of brane}
  Following the same picture as for entanglement entropy if we anchor an spacelike hypersurface on the brane which replace the asymptotic boundary of AdS and follow the volume conjecture of complexity which conjectures the quantum computational complexity of boundary field theory to be the the maximal volume of spacelike hypersurface
  \begin{equation}
    \mathcal{C}_{V}=\frac{V_{\Sigma}}{l\,G_{d+1}}
\end{equation}
The maximal volume for the special case when $t_{*}=0$ is given by, 
  \begin{eqnarray}
    V_{\Sigma}&=&2\int d\Omega_{d-1}\int_{0}^{r_{B}}dr~\sinh^{d-1}{r}\nonumber\\
    &=&\frac{4 \pi ^{d/2} \sinh ^d(r_{B} ) \, _2F_1\left(\frac{1}{2},\frac{d}{2};\frac{d+2}{2};-\sinh ^2(r_{B})\right)}{d~\Gamma \left(\frac{d}{2}\right)}
\end{eqnarray}
 the factor of two in front of integral is due to the two copies of AdS which are glued together at the brane position $r=r_{B}$. Comparing the maximal volume to the previous result without any brane insertion \eqref{eq: volume0} it can be seen that this is just twice of the latter. This feature of the bulk volume is same as was previously observed in \cite{Iwashita:2006zj} for the area functional in \eqref{eq:Sent_brane}. This indicates that the bulk geometry and therefore the volume does not change due to insertion of brane at the boundary, the doubling of the volume occurs only due to the two copies of AdS. Therefore following the volume conjecture the complexity is given by,
 \begin{align}
    C_{V}=&\frac{4 \pi ^{d/2} \sinh ^d(r_{B} ) \, _2F_1\left(\frac{1}{2},\frac{d}{2};\frac{d+2}{2};-\sinh ^2(r_{B})\right)}{G_{d+1}d~\Gamma \left(\frac{d}{2}\right)}
 \end{align}
 The result indicate the invariance of complexity of the boundary theory following the volume conjecture by insertion of the brane, except the factor of two.
 Therefore all the qualitative features of complexity like divergence structure and proportionality to the spatial volume of de Sitter, that we observed without brane are still valid in presence of brane. But, if we look the complexity from the perspective of the induced gravity theory living over de Sitter brane with the  effective Newton's constant \eqref{eq:eff_newt_const}, then additional dimension dependent numerically factors appear. 
 Same argument holds for $t_{*}\ne0$ numerically with the only distinction of factor of two due to the two copies of AdS, 
 \begin{eqnarray}\label{eq: ext_volume}
    V=2\int d\Omega_{d-1}\int dr (\sinh r\,\cosh t(r))^{d-1} \sqrt{1-\sinh ^2r\,t'(r)^2}
\end{eqnarray}
From here on, as the details remain same as was presented in section \ref{sec: Volume}, we refrain from presenting the expressions to avoid repetition. 
 

\subsection{Action Complexity in presence of brane}
Now we focus on the holographic complexity of the boundary theory sitting on de Sitter brane by following action conjecture which equate the quantum computational complexity of the boundary theory to the on shell action over the WDW patch anchored at a fixed boundary time. As the WDW patch is bounded by the null hypersurfaces in the bulk which are anchored at fixed boundary time the WDW patch geometry remains uneffected by the insertion of the brane at the boundary. But as we have discussed previously the full action now have a contribution from the brane,
\begin{align}
    S=&S_{\text{EH}}+S_{\text{brane}}\nonumber\\
    =&\frac{1}{16\pi G}\int d^{d+1}x\sqrt{-g}(R-2\Lambda)-\frac{1}{8\pi G}\int d^{d}x\sqrt{-h}~T.
\end{align}
But as we evaluate the action over the WDW patch which is anchored only at a fixed time at the boundary where the brane sit, the brane part of action do not contribute to the bulk part of the WDW action, therefore again the bulk part of WDW action remain uneffected by insersion of brane only a factor of two will be inserted to accommodate the two copies of bulk geometry. As has been discussed in details in section \ref{sec: action} the action over WDW patch also consist of other contributions, namely
\begin{itemize}
    \item \textbf{Boundary term:} Again as the geometry of null boundaries of WDW patch remains unchanged the boundary contribution to action which is evaluated at the null boundaries of WDW patch remains unchanged except the factor of two which account for doubling of spacetime.
    \item \textbf{Joint term:} As this contribution arises due to the intersection of two null boundaries it also remains invariant. There will be a factor of two multiplied again due to the $\mathbb{Z}_{2}$ symmetry across the brane the null vectors for both sides of null boundaries are equal and opposite to each other resulting in the addition of the respective contributions from both sides.
    \begin{align}
    I_{\text{joint}}=&-2\int d\Omega_{d-1}\sqrt{\gamma}\ln\left(-\frac{k^{+}.\bar{k}^{+}}{2}\right)-2\int d\Omega_{d-1}\sqrt{\gamma}\ln\left(-\frac{k^{-}.\bar{k}^{-}}{2}\right)\nonumber\\
    =&-4\int d\Omega_{d-1}\sqrt{\gamma}\ln\left(-\frac{k^{+}.\bar{k}^{+}}{2}\right)
    \end{align}
    \item \textbf{Counter term:} Again as this term is dependent on the geometry of null boundaries which remains invariant under the insertion of brane at the boundary this contribution also remains unchanged except the factor of two due to doubling of the bulk geometry.
\end{itemize}
Therefore the action complexity in presence of brane is given by,
\begin{eqnarray}
    C_{A}^{\text{brane}}=2C_{A}|_{\Lambda\rightarrow r_{B}}
\end{eqnarray}
The factor of two accounts for the two copies AdS spacetime glued at the brane position $r=r_{B}$. All the other qualitative features of complexity as the divergence structure or time-dependence of the complexity remains unchanged. 

This analysis exhibit that the insertion of an purely tensional brane at the boundary which is bound to backreact on the boundary geometry, do not effect the quantum computational complexity of the boundary theory. In other words our results indicates that the quantum computational complexity of boundary field theory living over an dynamical global de Sitter geometry is same as the quantum computational complexity of the field theory living over non-dynamical global de Sitter geometry.



\section{Discussion}\label{sec:conclusion}
In this work, we have carried out a study of quantum computational complexity for a quantum field theory living on a $d$-dimensional global de Sitter spacetime and a related set up of a field theory on a de Sitter brane embedded in AdS space. For the first case, our analysis is based on a holographic construction in which global $dS_d$ is realized as a slicing of an $(d+1)$-dimensional asymptotically AdS spacetime. This allows us to exploit the standard holographic prescriptions for complexity for the effective boundary theory lives on a time-dependent de Sitter background. Our setup provides a controlled and well-defined framework for investigating complexity in a genuinely cosmological and explicitly time-dependent setting.
A notable challenge in this construction arises from the explicit time dependence of the global de Sitter metric. This feature renders the determination of maximal bulk hypersurfaces nontrivial and significantly complicates the computation of holographic complexity, particularly within the complexity=volume (CV) conjecture. For the generic case of the maximal volume slices being anchored at a generic nonzero boundary time $t_\star$, we found that closed-form analytic expressions are generally inaccessible. In this case, we abandon analytic methods, and rely entirely on a numerical approach. An analytic result can be obtained only in a special and somewhat trivial case corresponding to the maximal surface anchored at $t=0$, for which the maximal volume takes the form
\begin{eqnarray}
V_{\Sigma}=\frac{2 \pi ^{d/2} \sinh ^d(\Lambda ) , 2F_1\left(\frac{1}{2},\frac{d}{2};\frac{d+2}{2};-\sinh ^2(\Lambda)\right)}{d~\Gamma \left(\frac{d}{2}\right)} , .
\end{eqnarray}
For generic boundary times $t_{*}$, however, the maximal surfaces must be constructed numerically. Our numerical analysis, presented in figures~\ref{fig:logVvsLambda} and \ref{fig:logVvsts}, reveals that the maximal volume and hence the volume complexity exhibits \emph{exponential dependence both on the UV cutoff $\Lambda$ and on the boundary anchoring time $t_\star$}: 
\begin{equation}
V_{\Sigma}\sim e^{(d-1)\Lambda}, \qquad V_{\Sigma}\sim e^{(d-1)t_{\star}} .
\end{equation}
This exponential growth is consistent with extensive character of complexity for local field theories: in local theories complexity is proportional to the number of degrees of freedom, i.e. the volume of constant global time spatial sections of de Sitter which is $(\sinh{\Lambda}\cosh{t_{*}})^{d-1}\sim e^{(d-1)t_\star}$. This exponential growth also indicates that the increase in complexity is due the increment in the degree of freedoms: we have put the CFT on an inflating background, and as a Hubble volume worth degrees of freedom are being added for every e-fold of inflation. This complexity growth is not due to the increase in the intrinsic quantum entanglement within the system with fixed degrees of freedom. Next, we find no evidence of \emph{hyperfast growth} in complexity, i.e. divergence of complexity at finite boundary time, of complexity within the CV proposal in global de Sitter spacetime. Such hyperfast complexity-growth phenomenon were observed previously for the case of static de Sitter patch holography \cite{Jorstad:2022mls}. This strongly points out that the hyperfast growth of complexity might be an observer dependent phenomenon.\\

Further, we also analyzed the complexity using the complexity=action (CA) prescription, where the complexity is identified with the on-shell gravitational action evaluated on the Wheeler De Witt (WDW) patch. In contrast to the volume-complexity, in this case, we were able to obtain an analytic expression for the complexity of a quantum conformal field theory living on $d$-dimensional global de Sitter spacetime. Although the full analytic expression is rather involved, nevertheless, several salient physical features can be easily extracted. First and foremost among these features is the UV-divergence structure: we found a universal scaling (with the UV-cutoff) depending on the spacetime dimension. For even $d$, the complexity behaves as
\begin{equation}
C_{\text{A}}\sim \frac{a}{\epsilon^{d-1}}+\frac{b}{\epsilon^{d+1}}+\cdots+\mathcal{O}(1),
\end{equation}
whereas for odd $d$ an additional logarithmic divergence appears,
\begin{equation}
C_{\text{A}}\sim \frac{a}{\epsilon^{d-1}}+\frac{b}{\epsilon^{d+1}}+\cdots +c\ln\epsilon+\ldots+\mathcal{O}(1).
\end{equation}
where $\Lambda$ is the bulk IR cutoff, and using the UV-IR relation \cite{Susskind:1998dq},  $\epsilon=e^{-\Lambda}$ is the boundary UV-cutoff, the lattice spacing. As expected, this leading order scaling for the UV-divergence identically to that of a local field theory in $d-1$ space dimensions, namely $C\sim \epsilon^{-(d-1)}$.

The next interesting feature is the time-dependence of the complexity, since we put the CFT on a time-dependent (inflating) background. The time dependence of the complexity extracted from the action exhibits exponential growth, consistent with the exponential expansion of the proper volume of spatial sections of the expoentially inflating de Sitter background:
\begin{align}
    C_{\text{A}}\sim e^{(d-1)t_{*}}
\end{align}
This dependence reinforces that the action-complexity, which is proportional to the number of boundary degrees of freedom, in a local theory scales with the spatial volume of the boundary (global de Sitter spacetime), i.e. .
\begin{align}
    C_{\text{A}}\sim  e^{(d-1)t_{*}} e^{(d-1)\Lambda}\sim (\sinh{\Lambda}\cosh{t_{*}})^{d-1}
\end{align}
Both these features of action-complexity, dependence on the bulk IR cutoff, and the time-dependence are identical to those of volume-complexity. Importantly, we again find no indication of hyperfast growth of action-complexity for CFT on global de Sitter! One of the main motivations for our project was to capture the entanglement structure, in particular complexity for field theories in time-dependent backgrounds exploiting the holographic correspondence. A key outcome of our analysis is that, in contrast to results obtained in the context of static patch holography where the boundary theory is conjectured to reside on the cosmological horizon and complexity has been argued to exhibit hyperfast growth we find no such behavior in global de Sitter. Instead, both the CV and CA prescriptions lead to exponential growth of complexity. This qualitative difference highlights the sensitivity of complexity to the choice of holographic framework and suggests that hyperfast growth may be a feature specific to horizon based descriptions rather than a universal property of de Sitter holography. Another potential origin of the hyperfast growth in the static patch holography studies could be the fact, that they include contributions due to the gravitational degrees of freedom in de Sitter (the stress tensor multiplet of the dual CFT on the stretched horizon captures the gravitational excitations or degrees of freedom in bulk of de Sitter in static coordinates), while in our set up only the physical effects of the nongravitational CFT degrees of freedom evolving in a passive/non-backreacting global de Sitter background has been considered.\\

In the second part of the paper, we also studied the effects of embedding a global de Sitter brane with finite tension near the asymptotic boundary of AdS. This is an interesting setup as the insertion of the brane give rise to an effective theory of gravity on the brane over which the boundary theory lives \cite{Bueno:2022log}. Therefore studying complexity in this setup will exhibit how complexity of field theory capture the effects of backreaction of brane tension over the background global de Sitter geometry. Our results indicates in this setup complexity following both volume and action conjecture show an increase (doubles) with respect to the complexity without any brane. This increment in complexity is also due to an increase in the bulk degree of freedom as the spacetime is now glued at the position of brane with an another copy of AdS which supports the non zero tension over the brane. Our results also makes explicit that the qualitative features of complexity like UV divergent structure and time dependence does not get effected by replacing the boundary by an brane with non-zero tension. Although we expect to see some non-trivial effects over complexity if brane is inserted at some bulk slice intersecting the boundary at a point. This can be achieved by inserting two branes rather than one as was considered in \cite{Aguilar-Gutierrez:2023tic,Fu:2024vin}. 
Another natural generalization is to include Einstein-Hilbert term into the brane action
\begin{align}
    S_{\text{brane}}=-\frac{1}{8\pi G}\int d^{d}x\sqrt{-h}~T+\frac{1}{16\pi G_{\text{brane}}}\int d^{d}x\sqrt{-h}\mathcal{R}_{\text{brane}}
\end{align}
We expect the presence of Einstein-Hilbert term to affect the complexity in a non-trivial way as the intersection of the two null surfaces of WDW patch which happens at the brane will give a non-zero contribution. Also if we consider fluctuating (backreacting) brane, then we might also see some nontrivial quantum gravity effects as observed in the case of AdS \cite{Fu:2024vin}.\\

There remain several open questions and interesting avenues of further investigations. In particular, it would be interesting to understand how notions such as the Lloyd bound should be reformulated, or possibly modified, in de Sitter spacetime where the number of degrees of freedom is exponentially increasing, i.e. does the exponential growth of complexity with time continue forever or is there some saturation level here as well arising from quantum mechanical uncertainty bounds.  Our universe is of course asymptotically de Sitter, it would be more realistic to consider holographic complexity for FLRW type spacetimes \cite{Rangamani:2015qga}. More broadly, extending our analysis to include black holes in the de Sitter boundary, i.e. studying a CFT at finite temperature (and rotation or finite chemical potential) in global de Sitter. That analysis will shed further light on the interplay between cosmological horizons in the boundary, black hole horizons in the bulk geometry, via the probe of quantum computational complexity. A further outstanding issue is use the dS/CFT correspondence \cite{Strominger:2001pn, Witten:2001kn, Maldacena:2002vr} directly by using schemes wherein quantum complexity is computed directly from the de Sitter Wheeler-de Witt wavefunction given by the appropriate dual CFT partition function \cite{Chakraborty:2023yed}. Yet another approach, which we plan to look at in a future follow-up work is exploiting the dS-dS set up \cite{Lewkowycz:2019xse} where the CFT dual to de Sitter space quantum gravity can be obtained from a well-defined $T\overline{T}$ deformation of the usual large $N$ CFT dual to quantum gravity in Anti de Sitter spacetime. For this exercise, one needs to study how holographic complexity behaves under holographic RG flows which are the bulk counterpart of $T\overline{T}$ deformation. This will also allow us to make contact with out previous work on complexity of little string theories \cite{Chakraborty:2020fpt,Katoch:2022hdf} obtained by $T\overline{T}$ deforming the worldsheet CFT in AdS. Another approach which is enormously promising for quantum theory in curved (dS) spacetimes is that using the von Neumann algebra of observables \cite{Chandrasekaran:2022cip, Liu:2025krl}. In \cite{Leutheusser:2025zvp} volume of spatial slices were interpreted in terms of the subalgebra based measure, namely the ``index of inclusion" which the authors have remarked bears close resemblance with complexity - one can pursue this connection further. Finally, one can consider holographic subregion complexity in de Sitter, for both spacelike and timelike subregions following the prescription outlined in \cite{Alishahiha:2025xml, Prihadi:2026nua}. In this context, it is interesting to study holographic dual of Krylov complexity\cite{Rabinovici:2023yex,Fu:2025kkh} in case of de Sitter spacetime, this has been recently studied for two dimensional de Sitter spacetime \cite{Heller:2025ddj}, while its generalization to higher dimensions remains an open problem.

\acknowledgments
We thank Arpan Bhattacharyya for helpful discussions. The work of SP and SR is supported by the CRG grant of the Anusandhan National Research
Foundation (ANRF), Department of Science and Technology (DST), CRG/2023/001120 (“Many facets
of complexity: From chaos to thermalization”). The work of SR is also supported by the IIT Hyderabad
seed grant SG/IITH/F171/2016-17/SG-47 as well the IITH funds RDF/IITH/F171/SR. 

\appendix
\section{Relation between AdS global time and dS global time}\label{sec: ads_ds_coord}
Let us consider global $AdS_{3}$,
\begin{equation}
    ds^{2}=d\rho^{2}-\cosh^{2}{\rho}~d\tau^{2}+\sinh^{2}{\rho}~d\theta^{2}
\end{equation}
embedded in $\mathbb{R}^{2,2}$, with
\begin{eqnarray}
    x^{0}&=&\cosh{\rho}\cos{\tau}\nonumber\\
    x^{1}&=&\sinh{\rho}\cos{\theta}\nonumber\\
    x^{2}&=&\sinh{\rho}\sin{\theta}\nonumber\\
    x^{3}&=&\cosh{\rho}\sin{\tau}
    \end{eqnarray}
    The global $dS_{2}$
    \begin{equation}
        ds^{2}=-dt^{2}+\cosh^{2}{t}d\phi^{2}
    \end{equation}
    is embedded in $\mathbb{R}^{2,1}$ as,
    \begin{eqnarray}
        x^{0}&=&\sinh{t}\nonumber\\
    x^{1}&=&\cosh{t}\cos{\phi}\nonumber\\
    x^{2}&=&\cosh{t}\sin{\phi}\nonumber\\
    \end{eqnarray}
If we choose an embedding, in $\mathbb{R}^{2,2}$ such as,
\begin{eqnarray}
    x^{0}&=&\sinh{t}\sinh{r}\nonumber\\
    x^{1}&=&\cosh{t}\cos{\phi}\sinh{r}\nonumber\\
    x^{2}&=&\cosh{t}\sin{\phi}\sinh{r}\nonumber\\ 
    x^{3}&=&\cosh{r}
\end{eqnarray}
then,
\begin{equation}
    ds^{2}=dr^{2}+\sinh^{2}{r}(-dt^{2}+\cosh^{2}{t}d\phi^{2})
\end{equation}
This identifies the global $AdS_{3}$ coordinates with global $dS_{2}$ coordinates,
\begin{eqnarray}\label{eq: adstods coord}
        \sinh{t}\sinh{r}&=&\cosh{\rho}\cos{\tau}\nonumber\\
    \cosh{t}\cos{\phi}\sinh{r}&=&\sinh{\rho}\cos{\theta}\nonumber\\
    \cosh{t}\sin{\phi}\sinh{r}&=&\sinh{\rho}\sin{\theta}\nonumber\\
    \cosh{r}&=&\cosh{\rho}\sin{\tau}\nonumber  
\end{eqnarray}
This relates the AdS golbal time $\tau$ with the dS golbal time $t$ as,
\begin{equation}
    \sinh{t}\tanh{r}=\cot{\tau}
\end{equation}
So a constant time slice in global AdS corresponds to,
\begin{eqnarray}
    t&=&\sinh^{-1}(\coth{r}\cot{\tau})\\
    t&=&\sinh^{-1}\left(\frac{\cosh{\rho}\cos{\tau}}{\sqrt{1+\cosh^{2}{\rho}\sin^{2}{\tau}}}\right)\\
\end{eqnarray}
Also constant $r$ surface corresponds to,
\begin{align}
    r=\cosh^{-1}(\cosh{\rho}\sin{\tau})
\end{align}

\section{Planar dS foliation of AdS}\label{sec: ads_ds_coord2}
If we choose the following embedding in $\mathbb{R}^{2,2}$
\begin{align}
    x^{0}&=\frac{R^2 \left(-T^2+X^2+1\right)+1}{2 R}\\
        x^{1}&=R X\\
    x^{2}&=\frac{R^2 \left(-T^2+X^2-1\right)+1}{2 R}\\
        x^{3}&=R T
\end{align}
Then the metric is,
\begin{align}
    ds^{2}=\frac{dR^{2}}{R^{2}}+R^{2}(-dT^{2}+dX^{2})
\end{align}
which describe the Poincare patch of AdS. If we perform a coordinate transformation
\begin{align}
    R&=\frac{\sinh (\rho )}{\eta },\\
    T&=\eta  \coth (\rho )
\end{align}
Then the resultant metric is,
\begin{align}
   ds^{2}= d\rho^2+\frac{\sinh ^{2}\rho  \left(-d\eta^2+dX^2\right)}{\eta ^2}
\end{align}
which describe the planer de Sitter foliation of Poincare patch of AdS spacetime.

These coordinates are related to the coordinates of global de Sitter foliation as,
\begin{align}
    \eta&=\frac{1}{\sin \theta  \cosh t+\sinh t}\\
    \rho&=r,\\
    X&=\frac{1}{\tanh t \sec \phi +\tan \phi }
\end{align}
The boundary $r\rightarrow\infty$ corresponds to $\rho\rightarrow\infty$ and at the boundary the relation between conformal(planer) time $\eta$ and global time $t$ is given by,
\begin{align}
  \eta&=\frac{1}{\cosh t \sin \phi -\sinh t}\\
\end{align}

\section{AdS foliation of AdS}\label{app: AdS_AdS}
In this appendix I will review the setup of \cite{Chen:2020uac} where they insert a brane in AdS$_{d+1}$ on which the geometry is AdS$_{d}$. Staring with the global AdS,
\begin{align}
    ds^{2}=-\cosh^{2}{\rho}~d\tau^{2}+d\rho^{2}+\sinh^{2}{\rho}(d\theta^{2}+\sin^{2}{\theta}d\Omega_{d-2}^{2})
\end{align}
by a coordinate transformation,
\begin{align}
    \tanh{\tilde{\rho}}= \tanh{\rho}\sin{\theta},\quad z= -2 \sinh{\rho} \cos{\theta}\pm 2 \sqrt{\sinh^{2}{\rho} \cos^{2}{\theta}+ 1} 
\end{align}
the metric can be recasted as,
\begin{align}
    ds^{2}=\frac{1}{z^{2}}\left(dz^{2}+\left(1+\frac{z^{2}}{4}\right)^{2}(-\cosh^{2}{\tilde{\rho}}~d\tau^{2}+d\tilde{\rho}^{2}+\sinh^{2}{\tilde{\rho}}d\Omega_{d-2}^{2})\right)
\end{align}
The brane is inserted at $z=z_{B}=0$ on which the geometry is $AdS_{d}$. 
\begin{align}
    z_{B}=-2 \sinh{r} \cos{\theta}\pm2 \sinh{r} \cos{\theta}
\end{align}
If we pick the plus sign above then the equality satisfies for $0<\theta<\frac{\pi}{2}$ and if we pick the minus sign then $\frac{\pi}{2}<\theta<\pi$. So from the global AdS perspective the brane lies to either upper or lower hemisphere of the AdS boundary.

\section{Useful Integrals}\label{app:integrals}
The integral,
\begin{eqnarray}
    \int dt \cosh^{2n}{t}&=&C^{n}_{2n}\frac{t}{2^{2n}}+\frac{1}{2^{2n-1}}\sum_{k=0}^{n-1}C^{k}_{2n}\frac{\sinh[2(n-k)t]}{2(n-k)}\\
    \int dt \cosh^{2n+1}{t}&=&\frac{1}{2^{2n}}\sum_{k=0}^{n}C^{k}_{2n+1}\frac{\sinh[(2n-2k+1){t}]}{2n-2k+1} 
\end{eqnarray}
\begin{align}
I_{1}=&\int_{0}^{\Lambda} dr~\sinh^{d}(r)\ln\left(\frac{\coth\left(\frac{r}{2}\right)}{\coth\left(\frac{\Lambda}{2}\right)}\right)\nonumber\\
=&\frac{2(1-\cosh (\Lambda )) \sinh ^{d-1}(\Lambda ) \text{sech}^{2 d}\left(\frac{\Lambda }{2}\right)}{(d+1)^2} \nonumber\\&\times\Bigg( (d+1) \ln \left(\tanh \left(\frac{\Lambda }{2}\right)\right) \, _2F_1\left(\frac{d+1}{2},d+1;\frac{d+3}{2};\tanh ^2\left(\frac{\Lambda }{2}\right)\right)\nonumber\\&- \, _3F_2\left(\frac{d}{2}+\frac{1}{2},\frac{d}{2}+\frac{1}{2},d+1;\frac{d}{2}+\frac{3}{2},\frac{d}{2}+\frac{3}{2};\tanh ^2\left(\frac{\Lambda }{2}\right)\right)\Bigg)\nonumber\\&-\frac{\sinh ^{d+1}(\Lambda ) \ln \left(\coth \left(\frac{\Lambda }{2}\right)\right) \, _2F_1\left(\frac{1}{2},\frac{d+1}{2};\frac{d+3}{2};-\sinh ^2(\Lambda )\right)}{d+1}
\end{align}
\begin{align}
I_{2}=&\int_{0}^{\Lambda} dr~(\sinh r)^{2d-2k-1}(\cosh r-1)^{-(d-2k-1)}\nonumber\\
    =&2^{d-k-1} \Gamma (k+1) \sinh ^{2 (d-k)}(\Lambda ) (\cosh (\Lambda )-1)^{-d+2 k+1} (\cosh (\Lambda )+1)^{k-d}\nonumber\\&\times \, _2\tilde{F}_1\left(k+1,-d+k+1;k+2;-\sinh ^2\left(\frac{\Lambda }{2}\right)\right)\nonumber\\
    I_{3}=&\int_{0}^{\Lambda} dr~(\sinh r)^{(2k+1)}(\cosh r-1)^{(d-2k-1)}\nonumber\\
    =&-\frac{1}{2} (-1)^{k-d} \Gamma (k+1) (\cosh (\Lambda )+1) \sinh ^{2 k}(\Lambda ) \sinh ^{2 (k-d)}\left(\frac{\Lambda }{2}\right) (\cosh (\Lambda )-1)^{d-2 k}\nonumber\\&\times \, _2\tilde{F}_1\left(k+1,-d+k+1;k+2;\cosh ^2\left(\frac{\Lambda }{2}\right)\right)+\frac{2^d (-1)^{(d-k)} \Gamma (k+1) \Gamma (d-k)}{\Gamma (d+1)}
\end{align}
\bibliographystyle{JHEP}
\bibliography{biblio.bib}

\providecommand{\href}[2]{#2}\begingroup\raggedright\begin{thebibliography}{10}

\bibitem{SupernovaCosmologyProject:1996grv}
{\scshape Supernova Cosmology Project} collaboration, \emph{{Measurements of the cosmological parameters Omega and Lambda from the first 7 supernovae at z{\ensuremath{>}}=0.35}}, \href{https://doi.org/10.1086/304265}{\emph{Astrophys. J.} {\bfseries 483} (1997) 565} [\href{https://arxiv.org/abs/astro-ph/9608192}{{\ttfamily astro-ph/9608192}}].

\bibitem{SupernovaSearchTeam:1998fmf}
{\scshape Supernova Search Team} collaboration, \emph{{Observational evidence from supernovae for an accelerating universe and a cosmological constant}}, \href{https://doi.org/10.1086/300499}{\emph{Astron. J.} {\bfseries 116} (1998) 1009} [\href{https://arxiv.org/abs/astro-ph/9805201}{{\ttfamily astro-ph/9805201}}].

\bibitem{Suzuki_2012}
N.~Suzuki, D.~Rubin, C.~Lidman, G.~Aldering, R.~Amanullah, K.~Barbary et~al., \emph{The hubble space telescope cluster supernova survey. v. improving the dark-energy constraints above z gt; 1 and building an early-type-hosted supernova sample*}, \href{https://doi.org/10.1088/0004-637X/746/1/85}{\emph{The Astrophysical Journal} {\bfseries 746} (2012) 85}.

\bibitem{Chernikov:1968zm}
N.A.~Chernikov and E.A.~Tagirov, \emph{{Quantum theory of scalar field in de Sitter space-time}}, {\emph{Ann. Inst. H. Poincare Phys. Theor. A} {\bfseries 9} (1968) 109}.

\bibitem{Mottola:1984ar}
E.~Mottola, \emph{{Particle Creation in de Sitter Space}}, \href{https://doi.org/10.1103/PhysRevD.31.754}{\emph{Phys. Rev. D} {\bfseries 31} (1985) 754}.

\bibitem{Allen:1985ux}
B.~Allen, \emph{{Vacuum States in de Sitter Space}}, \href{https://doi.org/10.1103/PhysRevD.32.3136}{\emph{Phys. Rev. D} {\bfseries 32} (1985) 3136}.

\bibitem{Schomblond:1976xc}
C.~Schomblond and P.~Spindel, \emph{{Unicity Conditions of the Scalar Field Propagator Delta(1) (x,y) in de Sitter Universe}}, {\emph{Ann. Inst. H. Poincare Phys. Theor.} {\bfseries 25} (1976) 67}.

\bibitem{Linde:1982uu}
A.D.~Linde, \emph{{Scalar Field Fluctuations in Expanding Universe and the New Inflationary Universe Scenario}}, \href{https://doi.org/10.1016/0370-2693(82)90293-3}{\emph{Phys. Lett. B} {\bfseries 116} (1982) 335}.

\bibitem{Vilenkin:1982wt}
A.~Vilenkin and L.H.~Ford, \emph{{Gravitational Effects upon Cosmological Phase Transitions}}, \href{https://doi.org/10.1103/PhysRevD.26.1231}{\emph{Phys. Rev. D} {\bfseries 26} (1982) 1231}.

\bibitem{Polarski:1991ek}
D.~Polarski, \emph{{Infrared divergences in de Sitter space}}, \href{https://doi.org/10.1103/PhysRevD.43.1892}{\emph{Phys. Rev. D} {\bfseries 43} (1991) 1892}.

\bibitem{Ford:1984hs}
L.H.~Ford, \emph{{Quantum Instability of De Sitter Space-time}}, \href{https://doi.org/10.1103/PhysRevD.31.710}{\emph{Phys. Rev. D} {\bfseries 31} (1985) 710}.

\bibitem{Tsamis:1996qq}
N.C.~Tsamis and R.P.~Woodard, \emph{{Quantum gravity slows inflation}}, \href{https://doi.org/10.1016/0550-3213(96)00246-5}{\emph{Nucl. Phys. B} {\bfseries 474} (1996) 235} [\href{https://arxiv.org/abs/hep-ph/9602315}{{\ttfamily hep-ph/9602315}}].

\bibitem{Starobinsky:1994bd}
A.A.~Starobinsky and J.~Yokoyama, \emph{{Equilibrium state of a selfinteracting scalar field in the De Sitter background}}, \href{https://doi.org/10.1103/PhysRevD.50.6357}{\emph{Phys. Rev. D} {\bfseries 50} (1994) 6357} [\href{https://arxiv.org/abs/astro-ph/9407016}{{\ttfamily astro-ph/9407016}}].

\bibitem{Burgess:2010dd}
C.P.~Burgess, R.~Holman, L.~Leblond and S.~Shandera, \emph{{Breakdown of Semiclassical Methods in de Sitter Space}}, \href{https://doi.org/10.1088/1475-7516/2010/10/017}{\emph{JCAP} {\bfseries 10} (2010) 017} [\href{https://arxiv.org/abs/1005.3551}{{\ttfamily 1005.3551}}].

\bibitem{Moreau:2018lmz}
G.~Moreau and J.~Serreau, \emph{{Stability of de Sitter spacetime against infrared quantum scalar field fluctuations}}, \href{https://doi.org/10.1103/PhysRevLett.122.011302}{\emph{Phys. Rev. Lett.} {\bfseries 122} (2019) 011302} [\href{https://arxiv.org/abs/1808.00338}{{\ttfamily 1808.00338}}].

\bibitem{Higuchi:2011vw}
A.~Higuchi, D.~Marolf and I.A.~Morrison, \emph{{de Sitter invariance of the dS graviton vacuum}}, \href{https://doi.org/10.1088/0264-9381/28/24/245012}{\emph{Class. Quant. Grav.} {\bfseries 28} (2011) 245012} [\href{https://arxiv.org/abs/1107.2712}{{\ttfamily 1107.2712}}].

\bibitem{Bernar:2014lna}
R.P.~Bernar, L.C.B.~Crispino and A.~Higuchi, \emph{{Infrared-finite graviton two-point function in static de Sitter space}}, \href{https://doi.org/10.1103/PhysRevD.90.024045}{\emph{Phys. Rev. D} {\bfseries 90} (2014) 024045} [\href{https://arxiv.org/abs/1405.3827}{{\ttfamily 1405.3827}}].

\bibitem{Hu:2018nxy}
B.-L.~Hu, \emph{{Infrared Behavior of Quantum Fields in Inflationary Cosmology -- Issues and Approaches: an overview}},  \href{https://arxiv.org/abs/1812.11851}{{\ttfamily 1812.11851}}.

\bibitem{Nielsen:2006cea}
M.A.~Nielsen, M.R.~Dowling, M.~Gu and A.C.~Doherty, \emph{{Quantum Computation as Geometry}}, \href{https://doi.org/10.1126/science.1121541}{\emph{Science} {\bfseries 311} (2006) 1133} [\href{https://arxiv.org/abs/quant-ph/0603161}{{\ttfamily quant-ph/0603161}}].

\bibitem{Jefferson:2017sdb}
R.~Jefferson and R.C.~Myers, \emph{{Circuit complexity in quantum field theory}}, \href{https://doi.org/10.1007/JHEP10(2017)107}{\emph{JHEP} {\bfseries 10} (2017) 107} [\href{https://arxiv.org/abs/1707.08570}{{\ttfamily 1707.08570}}].

\bibitem{Chagnet:2021uvi}
N.~Chagnet, S.~Chapman, J.~de~Boer and C.~Zukowski, \emph{{Complexity for Conformal Field Theories in General Dimensions}}, \href{https://doi.org/10.1103/PhysRevLett.128.051601}{\emph{Phys. Rev. Lett.} {\bfseries 128} (2022) 051601} [\href{https://arxiv.org/abs/2103.06920}{{\ttfamily 2103.06920}}].

\bibitem{Maldacena:1997re}
J.M.~Maldacena, \emph{{The Large $N$ limit of superconformal field theories and supergravity}}, \href{https://doi.org/10.4310/ATMP.1998.v2.n2.a1}{\emph{Adv. Theor. Math. Phys.} {\bfseries 2} (1998) 231} [\href{https://arxiv.org/abs/hep-th/9711200}{{\ttfamily hep-th/9711200}}].

\bibitem{Witten:1998qj}
E.~Witten, \emph{{Anti de Sitter space and holography}}, \href{https://doi.org/10.4310/ATMP.1998.v2.n2.a2}{\emph{Adv. Theor. Math. Phys.} {\bfseries 2} (1998) 253} [\href{https://arxiv.org/abs/hep-th/9802150}{{\ttfamily hep-th/9802150}}].

\bibitem{Ryu:2006bv}
S.~Ryu and T.~Takayanagi, \emph{{Holographic derivation of entanglement entropy from AdS/CFT}}, \href{https://doi.org/10.1103/PhysRevLett.96.181602}{\emph{Phys. Rev. Lett.} {\bfseries 96} (2006) 181602} [\href{https://arxiv.org/abs/hep-th/0603001}{{\ttfamily hep-th/0603001}}].

\bibitem{Ryu:2006ef}
S.~Ryu and T.~Takayanagi, \emph{{Aspects of Holographic Entanglement Entropy}}, \href{https://doi.org/10.1088/1126-6708/2006/08/045}{\emph{JHEP} {\bfseries 08} (2006) 045} [\href{https://arxiv.org/abs/hep-th/0605073}{{\ttfamily hep-th/0605073}}].

\bibitem{Susskind:2014moa}
L.~Susskind, \emph{{Entanglement is not enough}}, \href{https://doi.org/10.1002/prop.201500095}{\emph{Fortsch. Phys.} {\bfseries 64} (2016) 49} [\href{https://arxiv.org/abs/1411.0690}{{\ttfamily 1411.0690}}].

\bibitem{Belin:2021bga}
A.~Belin, R.C.~Myers, S.-M.~Ruan, G.~S{\'a}rosi and A.J.~Speranza, \emph{{Does Complexity Equal Anything?}}, \href{https://doi.org/10.1103/PhysRevLett.128.081602}{\emph{Phys. Rev. Lett.} {\bfseries 128} (2022) 081602} [\href{https://arxiv.org/abs/2111.02429}{{\ttfamily 2111.02429}}].

\bibitem{Belin:2022xmt}
A.~Belin, R.C.~Myers, S.-M.~Ruan, G.~S{\'a}rosi and A.J.~Speranza, \emph{{Complexity equals anything II}}, \href{https://doi.org/10.1007/JHEP01(2023)154}{\emph{JHEP} {\bfseries 01} (2023) 154} [\href{https://arxiv.org/abs/2210.09647}{{\ttfamily 2210.09647}}].

\bibitem{Myers:2024vve}
R.C.~Myers and S.-M.~Ruan, \emph{{Complexity Equals (Almost) Anything}},  3, 2024 [\href{https://arxiv.org/abs/2403.17475}{{\ttfamily 2403.17475}}].

\bibitem{Susskind:2014rva}
L.~Susskind, \emph{{Computational Complexity and Black Hole Horizons}}, \href{https://doi.org/10.1002/prop.201500092}{\emph{Fortsch. Phys.} {\bfseries 64} (2016) 24} [\href{https://arxiv.org/abs/1403.5695}{{\ttfamily 1403.5695}}].

\bibitem{Brown:2015lvg}
A.R.~Brown, D.A.~Roberts, L.~Susskind, B.~Swingle and Y.~Zhao, \emph{{Complexity, action, and black holes}}, \href{https://doi.org/10.1103/PhysRevD.93.086006}{\emph{Phys. Rev. D} {\bfseries 93} (2016) 086006} [\href{https://arxiv.org/abs/1512.04993}{{\ttfamily 1512.04993}}].

\bibitem{Brown:2015bva}
A.R.~Brown, D.A.~Roberts, L.~Susskind, B.~Swingle and Y.~Zhao, \emph{{Holographic Complexity Equals Bulk Action?}}, \href{https://doi.org/10.1103/PhysRevLett.116.191301}{\emph{Phys. Rev. Lett.} {\bfseries 116} (2016) 191301} [\href{https://arxiv.org/abs/1509.07876}{{\ttfamily 1509.07876}}].

\bibitem{Carmi:2016wjl}
D.~Carmi, R.C.~Myers and P.~Rath, \emph{{Comments on Holographic Complexity}}, \href{https://doi.org/10.1007/JHEP03(2017)118}{\emph{JHEP} {\bfseries 03} (2017) 118} [\href{https://arxiv.org/abs/1612.00433}{{\ttfamily 1612.00433}}].

\bibitem{Carmi:2017jqz}
D.~Carmi, S.~Chapman, H.~Marrochio, R.C.~Myers and S.~Sugishita, \emph{{On the Time Dependence of Holographic Complexity}}, \href{https://doi.org/10.1007/JHEP11(2017)188}{\emph{JHEP} {\bfseries 11} (2017) 188} [\href{https://arxiv.org/abs/1709.10184}{{\ttfamily 1709.10184}}].

\bibitem{Chapman:2016hwi}
S.~Chapman, H.~Marrochio and R.C.~Myers, \emph{{Complexity of Formation in Holography}}, \href{https://doi.org/10.1007/JHEP01(2017)062}{\emph{JHEP} {\bfseries 01} (2017) 062} [\href{https://arxiv.org/abs/1610.08063}{{\ttfamily 1610.08063}}].

\bibitem{Alishahiha:2015rta}
M.~Alishahiha, \emph{{Holographic Complexity}}, \href{https://doi.org/10.1103/PhysRevD.92.126009}{\emph{Phys. Rev. D} {\bfseries 92} (2015) 126009} [\href{https://arxiv.org/abs/1509.06614}{{\ttfamily 1509.06614}}].

\bibitem{Strominger:2001pn}
A.~Strominger, \emph{{The dS / CFT correspondence}}, \href{https://doi.org/10.1088/1126-6708/2001/10/034}{\emph{JHEP} {\bfseries 10} (2001) 034} [\href{https://arxiv.org/abs/hep-th/0106113}{{\ttfamily hep-th/0106113}}].

\bibitem{Banks:2005tn}
T.~Banks, L.~Mannelli and W.~Fischler, \emph{{Infrared divergences in dS/CFT}},  \href{https://arxiv.org/abs/hep-th/0507055}{{\ttfamily hep-th/0507055}}.

\bibitem{Balasubramanian:2002zh}
V.~Balasubramanian, J.~de~Boer and D.~Minic, \emph{{Notes on de Sitter space and holography}}, \href{https://doi.org/10.1016/S0003-4916(02)00020-9}{\emph{Class. Quant. Grav.} {\bfseries 19} (2002) 5655} [\href{https://arxiv.org/abs/hep-th/0207245}{{\ttfamily hep-th/0207245}}].

\bibitem{Susskind:2021omt}
L.~Susskind, \emph{{De Sitter Holography: Fluctuations, Anomalous Symmetry, and Wormholes}}, \href{https://doi.org/10.3390/universe7120464}{\emph{Universe} {\bfseries 7} (2021) 464} [\href{https://arxiv.org/abs/2106.03964}{{\ttfamily 2106.03964}}].

\bibitem{Jorstad:2022mls}
E.~J{\o}rstad, R.C.~Myers and S.-M.~Ruan, \emph{{Holographic complexity in dS$_{d+1}$}}, \href{https://doi.org/10.1007/JHEP05(2022)119}{\emph{JHEP} {\bfseries 05} (2022) 119} [\href{https://arxiv.org/abs/2202.10684}{{\ttfamily 2202.10684}}].

\bibitem{Susskind:2021esx}
L.~Susskind, \emph{{Entanglement and Chaos in De Sitter Space Holography: An SYK Example}}, \href{https://doi.org/10.22128/jhap.2021.455.1005}{\emph{JHAP} {\bfseries 1} (2021) 1} [\href{https://arxiv.org/abs/2109.14104}{{\ttfamily 2109.14104}}].

\bibitem{Alishahiha:2004md}
M.~Alishahiha, A.~Karch, E.~Silverstein and D.~Tong, \emph{{The dS/dS correspondence}}, \href{https://doi.org/10.1063/1.1848341}{\emph{AIP Conf. Proc.} {\bfseries 743} (2004) 393} [\href{https://arxiv.org/abs/hep-th/0407125}{{\ttfamily hep-th/0407125}}].

\bibitem{Gorbenko:2018oov}
V.~Gorbenko, E.~Silverstein and G.~Torroba, \emph{{dS/dS and $ T\overline{T} $}}, \href{https://doi.org/10.1007/JHEP03(2019)085}{\emph{JHEP} {\bfseries 03} (2019) 085} [\href{https://arxiv.org/abs/1811.07965}{{\ttfamily 1811.07965}}].

\bibitem{Lewkowycz:2019xse}
A.~Lewkowycz, J.~Liu, E.~Silverstein and G.~Torroba, \emph{{$ T\overline{T} $ and EE, with implications for (A)dS subregion encodings}}, \href{https://doi.org/10.1007/JHEP04(2020)152}{\emph{JHEP} {\bfseries 04} (2020) 152} [\href{https://arxiv.org/abs/1909.13808}{{\ttfamily 1909.13808}}].

\bibitem{Coleman:2021nor}
E.~Coleman, E.A.~Mazenc, V.~Shyam, E.~Silverstein, R.M.~Soni, G.~Torroba et~al., \emph{{De Sitter microstates from T$ \overline{T} $ + {\ensuremath{\Lambda}}$_{2}$ and the Hawking-Page transition}}, \href{https://doi.org/10.1007/JHEP07(2022)140}{\emph{JHEP} {\bfseries 07} (2022) 140} [\href{https://arxiv.org/abs/2110.14670}{{\ttfamily 2110.14670}}].

\bibitem{Batra:2024kjl}
G.~Batra, G.B.~De~Luca, E.~Silverstein, G.~Torroba and S.~Yang, \emph{{Bulk-local dS$_{3}$ holography: the matter with $ T\overline{T} $ + {\ensuremath{\Lambda}}$_{2}$}}, \href{https://doi.org/10.1007/JHEP10(2024)072}{\emph{JHEP} {\bfseries 10} (2024) 072} [\href{https://arxiv.org/abs/2403.01040}{{\ttfamily 2403.01040}}].

\bibitem{Hawking:2000da}
S.~Hawking, J.M.~Maldacena and A.~Strominger, \emph{{de Sitter entropy, quantum entanglement and AdS / CFT}}, \href{https://doi.org/10.1088/1126-6708/2001/05/001}{\emph{JHEP} {\bfseries 05} (2001) 001} [\href{https://arxiv.org/abs/hep-th/0002145}{{\ttfamily hep-th/0002145}}].

\bibitem{Randall:1999vf}
L.~Randall and R.~Sundrum, \emph{{An Alternative to compactification}}, \href{https://doi.org/10.1103/PhysRevLett.83.4690}{\emph{Phys. Rev. Lett.} {\bfseries 83} (1999) 4690} [\href{https://arxiv.org/abs/hep-th/9906064}{{\ttfamily hep-th/9906064}}].

\bibitem{Karch:2000ct}
A.~Karch and L.~Randall, \emph{{Locally localized gravity}}, \href{https://doi.org/10.1088/1126-6708/2001/05/008}{\emph{JHEP} {\bfseries 05} (2001) 008} [\href{https://arxiv.org/abs/hep-th/0011156}{{\ttfamily hep-th/0011156}}].

\bibitem{Emparan:2006ni}
R.~Emparan, \emph{{Black hole entropy as entanglement entropy: A Holographic derivation}}, \href{https://doi.org/10.1088/1126-6708/2006/06/012}{\emph{JHEP} {\bfseries 06} (2006) 012} [\href{https://arxiv.org/abs/hep-th/0603081}{{\ttfamily hep-th/0603081}}].

\bibitem{Emparan:2022ijy}
R.~Emparan, J.F.~Pedraza, A.~Svesko, M.~Toma{\v{s}}evi{\'c} and M.R.~Visser, \emph{{Black holes in dS$_{3}$}}, \href{https://doi.org/10.1007/JHEP11(2022)073}{\emph{JHEP} {\bfseries 11} (2022) 073} [\href{https://arxiv.org/abs/2207.03302}{{\ttfamily 2207.03302}}].

\bibitem{Almheiri:2019hni}
A.~Almheiri, R.~Mahajan, J.~Maldacena and Y.~Zhao, \emph{{The Page curve of Hawking radiation from semiclassical geometry}}, \href{https://doi.org/10.1007/JHEP03(2020)149}{\emph{JHEP} {\bfseries 03} (2020) 149} [\href{https://arxiv.org/abs/1908.10996}{{\ttfamily 1908.10996}}].

\bibitem{Chen:2020uac}
H.Z.~Chen, R.C.~Myers, D.~Neuenfeld, I.A.~Reyes and J.~Sandor, \emph{{Quantum Extremal Islands Made Easy, Part I: Entanglement on the Brane}}, \href{https://doi.org/10.1007/JHEP10(2020)166}{\emph{JHEP} {\bfseries 10} (2020) 166} [\href{https://arxiv.org/abs/2006.04851}{{\ttfamily 2006.04851}}].

\bibitem{Emparan:1999fd}
R.~Emparan, G.T.~Horowitz and R.C.~Myers, \emph{{Exact description of black holes on branes. 2. Comparison with BTZ black holes and black strings}}, \href{https://doi.org/10.1088/1126-6708/2000/01/021}{\emph{JHEP} {\bfseries 01} (2000) 021} [\href{https://arxiv.org/abs/hep-th/9912135}{{\ttfamily hep-th/9912135}}].

\bibitem{Emparan:1999wa}
R.~Emparan, G.T.~Horowitz and R.C.~Myers, \emph{{Exact description of black holes on branes}}, \href{https://doi.org/10.1088/1126-6708/2000/01/007}{\emph{JHEP} {\bfseries 01} (2000) 007} [\href{https://arxiv.org/abs/hep-th/9911043}{{\ttfamily hep-th/9911043}}].

\bibitem{Emparan:2002px}
R.~Emparan, A.~Fabbri and N.~Kaloper, \emph{{Quantum black holes as holograms in AdS brane worlds}}, \href{https://doi.org/10.1088/1126-6708/2002/08/043}{\emph{JHEP} {\bfseries 08} (2002) 043} [\href{https://arxiv.org/abs/hep-th/0206155}{{\ttfamily hep-th/0206155}}].

\bibitem{Iwashita:2006zj}
Y.~Iwashita, T.~Kobayashi, T.~Shiromizu and H.~Yoshino, \emph{{Holographic entanglement entropy of de Sitter braneworld}}, \href{https://doi.org/10.1103/PhysRevD.74.064027}{\emph{Phys. Rev. D} {\bfseries 74} (2006) 064027} [\href{https://arxiv.org/abs/hep-th/0606027}{{\ttfamily hep-th/0606027}}].

\bibitem{Kushihara:2021fbr}
K.~Kushihara, K.~Izumi and T.~Shiromizu, \emph{{Holographic entanglement entropy of a de Sitter braneworld with Lovelock terms}}, \href{https://doi.org/10.1093/ptep/ptab038}{\emph{PTEP} {\bfseries 2021} (2021) 043E01} [\href{https://arxiv.org/abs/2102.12597}{{\ttfamily 2102.12597}}].

\bibitem{Reynolds_2017}
A.P.~Reynolds and S.F.~Ross, \emph{Complexity in de sitter space}, \href{https://doi.org/10.1088/1361-6382/aa8122}{\emph{Classical and Quantum Gravity} {\bfseries 34} (2017) 175013}.

\bibitem{Susskind:1998dq}
L.~Susskind and E.~Witten, \emph{{The Holographic bound in anti-de Sitter space}},  \href{https://arxiv.org/abs/hep-th/9805114}{{\ttfamily hep-th/9805114}}.

\bibitem{Lloyd:2000cry}
S.~Lloyd, \emph{{Ultimate physical limits to computation}}, \href{https://doi.org/10.1038/35023282}{\emph{Nature} {\bfseries 406} (2000) 1047} [\href{https://arxiv.org/abs/quant-ph/9908043}{{\ttfamily quant-ph/9908043}}].

\bibitem{Lehner:2016vdi}
L.~Lehner, R.C.~Myers, E.~Poisson and R.D.~Sorkin, \emph{{Gravitational action with null boundaries}}, \href{https://doi.org/10.1103/PhysRevD.94.084046}{\emph{Phys. Rev. D} {\bfseries 94} (2016) 084046} [\href{https://arxiv.org/abs/1609.00207}{{\ttfamily 1609.00207}}].

\bibitem{Reynolds:2016rvl}
A.~Reynolds and S.F.~Ross, \emph{{Divergences in Holographic Complexity}}, \href{https://doi.org/10.1088/1361-6382/aa6925}{\emph{Class. Quant. Grav.} {\bfseries 34} (2017) 105004} [\href{https://arxiv.org/abs/1612.05439}{{\ttfamily 1612.05439}}].

\bibitem{deHaro:2000wj}
S.~de~Haro, K.~Skenderis and S.N.~Solodukhin, \emph{{Gravity in warped compactifications and the holographic stress tensor}}, \href{https://doi.org/10.1088/0264-9381/18/16/307}{\emph{Class. Quant. Grav.} {\bfseries 18} (2001) 3171} [\href{https://arxiv.org/abs/hep-th/0011230}{{\ttfamily hep-th/0011230}}].

\bibitem{Bueno:2022log}
P.~Bueno, R.~Emparan and Q.~Llorens, \emph{{Higher-curvature gravities from braneworlds and the holographic c-theorem}}, \href{https://doi.org/10.1103/PhysRevD.106.044012}{\emph{Phys. Rev. D} {\bfseries 106} (2022) 044012} [\href{https://arxiv.org/abs/2204.13421}{{\ttfamily 2204.13421}}].

\bibitem{Nojiri:2002wn}
S.~Nojiri and S.D.~Odintsov, \emph{{Newton potential in deSitter brane world}}, \href{https://doi.org/10.1016/S0370-2693(02)02859-9}{\emph{Phys. Lett. B} {\bfseries 548} (2002) 215} [\href{https://arxiv.org/abs/hep-th/0209066}{{\ttfamily hep-th/0209066}}].

\bibitem{Calcagni:2005vn}
G.~Calcagni, \emph{{de Sitter thermodynamics and the braneworld}}, \href{https://doi.org/10.1088/1126-6708/2005/09/060}{\emph{JHEP} {\bfseries 09} (2005) 060} [\href{https://arxiv.org/abs/hep-th/0507125}{{\ttfamily hep-th/0507125}}].

\bibitem{Panella:2023lsi}
E.~Panella and A.~Svesko, \emph{{Quantum Kerr-de Sitter black holes in three dimensions}}, \href{https://doi.org/10.1007/JHEP06(2023)127}{\emph{JHEP} {\bfseries 06} (2023) 127} [\href{https://arxiv.org/abs/2303.08845}{{\ttfamily 2303.08845}}].

\bibitem{Myers:2013lva}
R.C.~Myers, R.~Pourhasan and M.~Smolkin, \emph{{On Spacetime Entanglement}}, \href{https://doi.org/10.1007/JHEP06(2013)013}{\emph{JHEP} {\bfseries 06} (2013) 013} [\href{https://arxiv.org/abs/1304.2030}{{\ttfamily 1304.2030}}].

\bibitem{Aguilar-Gutierrez:2023tic}
S.E.~Aguilar-Gutierrez, A.K.~Patra and J.F.~Pedraza, \emph{{Entangled universes in dS wedge holography}}, \href{https://doi.org/10.1007/JHEP10(2023)156}{\emph{JHEP} {\bfseries 10} (2023) 156} [\href{https://arxiv.org/abs/2308.05666}{{\ttfamily 2308.05666}}].

\bibitem{Fu:2024vin}
Y.~Fu and K.-Y.~Kim, \emph{{Wedge holographic complexity in Karch-Randall braneworld}}, \href{https://doi.org/10.1007/JHEP01(2025)174}{\emph{JHEP} {\bfseries 01} (2025) 174} [\href{https://arxiv.org/abs/2412.00852}{{\ttfamily 2412.00852}}].

\bibitem{Rangamani:2015qga}
M.~Rangamani, M.~Rozali and M.~Van~Raamsdonk, \emph{{Cosmological Particle Production at Strong Coupling}}, \href{https://doi.org/10.1007/JHEP09(2015)213}{\emph{JHEP} {\bfseries 09} (2015) 213} [\href{https://arxiv.org/abs/1505.03901}{{\ttfamily 1505.03901}}].

\bibitem{Witten:2001kn}
E.~Witten, \emph{{Quantum gravity in de Sitter space}},  in \emph{{Strings 2001: International Conference}}, 6, 2001 [\href{https://arxiv.org/abs/hep-th/0106109}{{\ttfamily hep-th/0106109}}].

\bibitem{Maldacena:2002vr}
J.M.~Maldacena, \emph{{Non-Gaussian features of primordial fluctuations in single field inflationary models}}, \href{https://doi.org/10.1088/1126-6708/2003/05/013}{\emph{JHEP} {\bfseries 05} (2003) 013} [\href{https://arxiv.org/abs/astro-ph/0210603}{{\ttfamily astro-ph/0210603}}].

\bibitem{Chakraborty:2023yed}
T.~Chakraborty, J.~Chakravarty, V.~Godet, P.~Paul and S.~Raju, \emph{{The Hilbert space of de Sitter quantum gravity}}, \href{https://doi.org/10.1007/JHEP01(2024)132}{\emph{JHEP} {\bfseries 01} (2024) 132} [\href{https://arxiv.org/abs/2303.16315}{{\ttfamily 2303.16315}}].

\bibitem{Chakraborty:2020fpt}
S.~Chakraborty, G.~Katoch and S.R.~Roy, \emph{{Holographic complexity of LST and single trace $ T\overline{T} $}}, \href{https://doi.org/10.1007/JHEP03(2021)275}{\emph{JHEP} {\bfseries 03} (2021) 275} [\href{https://arxiv.org/abs/2012.11644}{{\ttfamily 2012.11644}}].

\bibitem{Katoch:2022hdf}
G.~Katoch, S.~Mitra and S.R.~Roy, \emph{{Holographic complexity of LST and single trace $ T\overline{T} $, $ J\overline{T} $ and $ T\overline{J} $ deformations}}, \href{https://doi.org/10.1007/JHEP10(2022)143}{\emph{JHEP} {\bfseries 10} (2022) 143} [\href{https://arxiv.org/abs/2208.02314}{{\ttfamily 2208.02314}}].

\bibitem{Chandrasekaran:2022cip}
V.~Chandrasekaran, R.~Longo, G.~Penington and E.~Witten, \emph{{An algebra of observables for de Sitter space}}, \href{https://doi.org/10.1007/JHEP02(2023)082}{\emph{JHEP} {\bfseries 02} (2023) 082} [\href{https://arxiv.org/abs/2206.10780}{{\ttfamily 2206.10780}}].

\bibitem{Liu:2025krl}
H.~Liu, \emph{{Lectures on entanglement, von Neumann algebras, and emergence of spacetime}},  in \emph{{Theoretical Advanced Study Institute in Elementary Particle Physics 2023}: {Aspects of Symmetry}}, 10, 2025 [\href{https://arxiv.org/abs/2510.07017}{{\ttfamily 2510.07017}}].

\bibitem{Leutheusser:2025zvp}
S.~Leutheusser and H.~Liu, \emph{{Volume as an index of a subalgebra}},  \href{https://arxiv.org/abs/2508.00056}{{\ttfamily 2508.00056}}.

\bibitem{Alishahiha:2025xml}
M.~Alishahiha, \emph{{Timelike Holographic Complexity}},  \href{https://arxiv.org/abs/2510.25700}{{\ttfamily 2510.25700}}.

\bibitem{Prihadi:2026nua}
H.L.~Prihadi, M.A.R.~Al-Faritsi, R.R.~Firdaus, F.~Khairunnisa, Y.P.~Sarwono and F.P.~Zen, \emph{{Holographic timelike entanglement and subregion complexity with scalar hair}}, \href{https://doi.org/10.1007/JHEP04(2026)174}{\emph{JHEP} {\bfseries 04} (2026) 174} [\href{https://arxiv.org/abs/2601.18310}{{\ttfamily 2601.18310}}].

\bibitem{Rabinovici:2023yex}
E.~Rabinovici, A.~S{\'a}nchez-Garrido, R.~Shir and J.~Sonner, \emph{{A bulk manifestation of Krylov complexity}}, \href{https://doi.org/10.1007/JHEP08(2023)213}{\emph{JHEP} {\bfseries 08} (2023) 213} [\href{https://arxiv.org/abs/2305.04355}{{\ttfamily 2305.04355}}].

\bibitem{Fu:2025kkh}
Y.~Fu, H.-S.~Jeong, K.-Y.~Kim and J.F.~Pedraza, \emph{{Toward Krylov-based holography in double-scaled SYK}},  \href{https://arxiv.org/abs/2510.22658}{{\ttfamily 2510.22658}}.

\bibitem{Heller:2025ddj}
M.P.~Heller, F.~Ori, J.~Papalini, T.~Schuhmann and M.-T.~Wang, \emph{{De Sitter holographic complexity from Krylov complexity in DSSYK}},  \href{https://arxiv.org/abs/2510.13986}{{\ttfamily 2510.13986}}.

\end{thebibliography}\endgroup




\end{document}